\documentclass[reprint,eqsecnum,floats,aps,amsmath,amssymb,nofootinbib,prd,onecolumn, showpacs]{revtex4-2}

\usepackage{bm}
\usepackage{graphicx,physics}
\usepackage{graphicx}
\usepackage{amsmath,amssymb,mathtools,mathrsfs}
\usepackage{amsmath,amssymb,mathrsfs}
\usepackage{hyperref}
\usepackage{graphicx}
\usepackage{arydshln}
\usepackage{xcolor}
\usepackage{braket}
\usepackage{tensor}
\usepackage{enumitem,array,textcomp}
\usepackage[scr=boondoxo]{mathalpha}
\usepackage{caption}
\usepackage{subcaption}
\usepackage{tikz}

\usepackage{color}

\begin{document}

\title{Expansion-sensitive coupling of a local quantum system in de Sitter cosmology}

\author{Jorma Louko}
\email{Electronic address: jorma.louko@nottingham.ac.uk}
\affiliation{School of Mathematical Sciences, University of Nottingham, Nottingham NG7 2RD, UK}

\author{Guillermo A. Mena Marug\'an}
\email{Electronic address: mena@iem.cfmac.csic.es}
\affiliation{Instituto de Estructura de la Materia, IEM-CSIC, C/ Serrano 121, 28006 Madrid, Spain}

\author{Alvaro Torres-Caballeros}
\email{Electronic address: alvaro.torres@iem.cfmac.csic.es}
\affiliation{School of Mathematical Sciences, University of Nottingham, Nottingham NG7 2RD, UK}
\affiliation{Instituto de Estructura de la Materia, IEM-CSIC, C/ Serrano 121, 28006 Madrid, Spain}

\date{September 2025. Revised November 2025.\\ aaPublished in Phys.\ Rev.\ D \textbf{112}, 125024 (2025), doi.org/10.1103/5twj-hl9b.}

\def\l{\hat{l}}
\def\f{\mathring{\phi}_{nlm}}
\def\p{\mathring{\Pi}_{nlm}}
\def\a{\mathring{a}_{nlm}}
\def\aa{\mathring{a}_{nlm}^{*}}
\def\z{z_{k\l}}
\def\bl{\bar{l}}
\def\bll{\hat{\bar{l}}}
\def\bk{\Bar{k}}
\def\ii{\text{int}}
\def\ee{\text{ext}}
\def\tP{\Tilde{P}}
\def\tQ{\Tilde{Q}}
\def\tA{\Tilde{A}}
\def\b{b_{\l}}
\def\x{\xi_{nl}}
\def\f{f_{nl}}
\def\g{g_{nl}}
\def\vaf{\vartheta_{nl}^{(f)}}
\def\vag{\vartheta_{nl}^{(g)}}
\def\s{s_{\l}}
\def\h{h_{nl}}
\def\bt{\Tilde{b}_{\l}}
\def\Hb{\mathscr{H}}
\newcommand{\dif}[1]{\overset{*}{#1}}
\newcommand{\ddif}[1]{\overset{**}{#1}}
\def\upr{\underline{\Pi}^{rr}}
\def\upt{\underline{\Pi}^{\theta \theta}}
\def\upf{\underline{\Pi}_{\Phi_0}}
\def\uc{\underline{c}}
\def\up{\underline{p_b}}
\def\bN{\Bar{N}}
\def\ti{\tau_{\text{int}}}
\def\fa{\left[ 1-\l^2 \right]}
\def\ga{\gamma_{\l_0}}
\def\PT{\mathscr{T}}
\def\modes{\text{osc}}

\begin{abstract}
When a local quantum system couples to a quantum field in a cosmological spacetime, the time dependence of the coupling strength is conventionally taken to reflect the design of the local quantum system but not to depend on the large-scale structure of the universe. In this paper, we consider a novel coupling that incorporates additional time dependence that reflects the cosmological expansion, as motivated by structures that appear in quantum cosmology. We focus on a conformal scalar field in a de Sitter Friedmann-Lema\^{\i}tre-Robertson-Walker cosmology with flat but compact spatial sections in $3+1$ dimensions, and a comoving Unruh-DeWitt detector: the novel coupling posits the detector to couple to the scaled scalar field that appears in the conformally related static spacetime. We survey the differences between the conventional and novel coupling, for detectors that couple to the full field and detectors that couple only to specific field modes, and for detectors with proper time internal dynamics and detectors with conformal time internal dynamics. We also briefly discuss noncompact spatial sections and single-mode detectors with discontinuous time dependence. We find that the novel coupling tends to enhance the de-excitation peaks in the detector's response, particularly for single-mode detectors.
\end{abstract}

\maketitle

\section{Introduction}\label{sec:intro}

Among the fundamental questions that the generalisation of (Minkowskian) quantum field theory to curved spacetimes has brought into light are the concepts of vacuum and particles. For instance, in curved spacetimes, the number of particles in a state is observer-dependent \cite{BD, Crispino} and the particle formulation of the theory may lead to ambiguities that univocally are traced to the separation of the space of solutions into positive- and negative-frequencies~\cite{Carroll}. Some results and remarks indicate taking the field observables, such as correlation functions, as more fundamental objects rather than a description of quantum states in terms of particles~\cite{Mukhanov}; and, although the notion of a particle is a topic that has been studied on its own, for instance in Refs.~\cite{Fulling,  Parker}, the point of view that we shall adopt is of addressing this notion to remain purely mathematical until a method of observation is specified \cite{BD, Carroll, Takagi}. Hence, it is common practice to introduce the (idealised) concept of a \textit{particle detector}, defined as a point-like object modelled as a field that is characterised by energy levels and is coupled linearly to a given field via a \textit{monopole moment}. Although this so-called Unruh–DeWitt particle detector is an idealised model distinct from detectors used in high-energy physics (e.g. in particle accelerators), the pointlike two-level system captures conceptual features at a fundamental level~\cite{Carroll}. As introduced, for example, by Refs.~\cite{BD,Takagi}, this particle detector can be roughly seen as a collection of \textit{atoms\/} interacting in a very specific form with a field in a given spacetime along a particular trajectory. The interaction will have an associated probability of these atoms being excited and de-exited (intuitively associated with the detector probing particles), and that can be factorised into a part regarding the internal structure of the system times (and hence the probability is directly proportional to) another that solely depends on the field and its interaction with the detector, the response function. We shall focus our attention on the response function since it is the one with physical significance (for a detailed discussion of the conceptual foundations of the Unruh–DeWitt detector, readers are referred to Refs. \cite{Unruh:1976db,DeWitt:1980hx}). These detectors are usually composed of a \textit{switching} function, which serves as a window to dictate the duration of the interaction, an energy gap term, which depends on the difference between the detector's energy levels, 
and the Wightman function (defined for the case of a scalar field as $\bra{0} \hat{\phi}(t,\vec{x}), \hat{\phi}(t',\vec{x}') \ket{0}$) pulled back to the detector's trajectory. Moreover, it is usual to define the detector so that it can measure its proper time from which it will be able to define (if symmetries allow) the corresponding positive- and negative-frequency~\cite{Carroll, Takagi}.

In an expanding cosmology, an additional consideration must be addressed, namely, that spacetime is evolving. Hence, it is reasonable to think that part of the field evolution could be reassigned to a part corresponding to the evolution of the background itself, modifying in this manner the usual notion addressed to the field dynamics. One can trace the origins of this idea to cosmology, especially in the study of cosmological perturbations on spacetimes with compact spatial topology, where time-dependent rescalings of fields are employed to partially absorb the evolution of the cosmological background \cite{HandH, PerturbMukhanov, GaugePerturb, ScalingMukhanov, Mik, JGJV}. 

Let us briefly develop the idea and introduce the concept we refer to as dynamics. For the sake of the example, consider a scalar field $\phi$ in a globally hyperbolic, spatially homogeneous and isotropic spacetime. We begin by describing the system at time equal $t_0$ by the corresponding Fourier coefficients and Fourier modes and promote these coefficients to operators living in Fock space, or Hilbert space in the case of one particle. Note, however, that if we choose to describe the system at time $t\ne t_0$ instead, different (or not necessarily the same) Fourier coefficients are obtained. We can choose to work with the same Fock space as previously and note that both field descriptions are related by a Bogoliubov transformation. Hence, by choosing the same representation for both descriptions of the field, this latter transformation is found by solving the corresponding equations of motion, assuming the initial conditions of the system are well-posed. We can thus understand this Bogoliubov transformation as dynamics. Moreover, an important observation is to be made. Indeed, to simplify the equations of motion, in cosmological spacetimes it is convenient to introduce what is known as the conformal field~$\chi$, equal (or proportional) to the scalar field times the scale factor, and even though the ground states of the two fields correspond to the same vacuum, their respective dynamics are different. This fact usually does not pose any concern, since 
the dynamics of the two fields are closely related; i.e.\ by knowing the solution $\chi$ at time equal $t_0$, we also know $\phi$ at time equal~$t_0$. Nevertheless, it is noteworthy that, in the case of a massless field, the action of the conformal field can be recast as that of a field in flat spacetime, thus decoupled from geometrical dynamical contributions. Similarly, in the case of a massive scalar field, we arrive at the same results for the kinetic term in the Lagrangian of the conformal field, alongside acquiring an effective mass that depends on the scale factor. In fact, one of the primary motivations of introducing the conformal field is to be able to identify a canonical kinetic\footnote{We refer to a canonical term $X$ for a scalar field $\phi$ as the term defined by $\tfrac12\bigl(\dot{\phi}^2 - (\partial_i \phi)^2\bigr)$, where the overdot denotes a time derivative and $\partial_i$ the space derivatives. The canonical scalar field corresponds then to the choice of a densitized Lagrangian $\mathcal{L}= X-V(\phi)$, where $V(\phi)$ is the potential of the field~\cite{Weinberg}.} term of the theory~\cite{conformal}. Therefore, one could think of the dynamics of the conformal field, governed by the corresponding canonical kinetic term, as the \textit{proper\/} dynamics of the field\footnote{For an argument based on unitary implementable dynamics, see Ref.~\cite{Triada}.}.

The study of quantum field theory in curved spacetimes, such as in cosmological backgrounds, covers a wide range of areas, some of which have in fact been very actively explored recently. One active direction concerns the characterisation and harvesting of entanglement in quantum fields \cite{AgulloBonga2, EduMM2, EduMM1}, with some works exploring time-dependent detector couplings as a tool to manage causal separation and finite-time interactions, e.g.\ Refs.~\cite{Valentini:1991eah, Reznik, RRS, Kempf, HHMSZ, ZhangYu, FOZ, SurMannCong,  LML}. Another notable line of work focuses on the unitary evolution of quantum states in dynamical geometries. In particular, several works argue that the unitary implementability of field dynamics provides a criterion for selecting a unique quantum representation (up to unitary equivalence) \cite{Tetrada,Triada,Triada+2,Triada1} and argue that it is important for defining a consistent relationship between the Schr\"odinger and Heisenberg pictures in cosmological and black-hole settings~\cite{Giddings}. These advances provide a strong motivation for the present work, as both perspectives ---time-dependent couplings and unitary implementable field dynamics--- shed light on different aspects of quantum field evolution in dynamical backgrounds, and may prove relevant for pre-inflationary analyses such as that presented in Ref.~\cite{Conroy}.

Therefore, turning back to the context of detectors, and given the growing interest in this field, it becomes natural to investigate the effects of introducing a time-dependent scaling in the detector's coupling. In previous approaches in this area, this time dependence has been introduced by hand, e.g.\ to model detectors sensitive to the Berry phase rather than excitation responses~\cite{Quach:2021vzo}, to recover the long-interaction limit~\cite{UnruhWaiting,Parry:2025wub}, 
or to account for finite-time interactions motivated by physical considerations such as distinguishing early- and late-time behaviour \cite{Echo, LoukoMann, BlackHoleFall}. In contrast, following our previous discussion, we will consider here a background-dependent coupling, and in particular, the coupling to the conformal field, since mathematically it is equivalent to considering the usual coupling to the field $\phi$ with a background-dependent and hence time-dependent scaling in that coupling. More specifically, we will study the different characteristics of the particle measuring in de Sitter spacetime resulting from these two fields with distinct dynamics ---one with standard dynamics and the other with \textit{proper\/} dynamics. In this way, we will determine which corresponding detector reading better aligns with our intuitive notion of a particle. Moreover, our requirement that the time dependence of the detector coupling be in a direct functional form with respect to the spacetime allows that it can be fully specified by endowing the detector with the capacity to perform a direct reading of the background spacetime without any knowledge in advance of its behaviour. 
In essence, for a time dependence of the coupling restricted to come entirely from the dynamical background, we could think of a detector device able to measure this background simultaneously and independently, and then engineer or design a data processing system that incorporates this information into the final signal by means of a Fourier convolution. This convolution would have the same effect as if the original detector's response function were multiplied by the background dependent function of our choice. In total, the set-up would modify the detector's outcome in terms of the background signal without a prior knowledge of how this background evolves or the use of any externally prescribed control profile required to emulate its evolution. Hence, in this sense,  through these independent readings of spacetime, the new detector under consideration will be capable of absorbing the possible excitations caused by the background dynamics and be well-adapted to the expanding cosmology in the considered case of de Sitter spacetime.

To achieve these objectives, 
we consider two detector models within a highly symmetric but curved spacetime in which the quantised field has solutions in terms of known functions: the de Sitter cosmology~\cite{BAllen}. The first detector model is a conventional one, according to the literature, serving us as our \textit{control group\/} for comparison purposes and is referred to as the \textit{standard\/} detector along the work. For the second one, we exploit the subtle freedom in the construction of particle detectors and introduce a background-dependent scaling in the detector coupling, calling this the \textit{novel\/} detector.  Finally, we expose both particle detectors to different conditions including long duration intervals, a coupling with respect to a different reparametrized time, ideal switches and a modification of the topology of the spacetime. This allows us to compare their respective different characteristics and highlight that the resulting measurements with the novel coupling are closely related to those in a static spacetime, providing evidence that the new detector is well-suited to the considered spacetime. 

The structure of the paper is as follows.

In Sec.\ \ref{sec:fieldandspacetime} we develop quantum field theory for a massive scalar field minimally coupled to a locally de Sitter background in $3+1$ dimensions. 
We work in the spatially flat Friedmann-Lema\^{\i}tre-Robertson-Walker (FLRW) foliation and choose the spatial hypersurfaces to have the compact topology of the three-torus. We 
choose the Bunch-Davies vacuum for the ground state of our system~\cite{Mukhanov,Bunch:1978yq}, and we select a specific scalar field mass value to simplify the computations, such that the time-dependent modes are expressible as elementary functions. Moreover, for comparison purposes (which is our principal objective in the present article), we discard the zero mode, which would need special treatment. 

In Sec.\ \ref{sec:UDWdetector} we introduce the Unruh-DeWitt (pointlike) detectors~\cite{Unruh:1976db,DeWitt:1980hx}, 
for comoving trajectories in the expanding locally de Sitter metric. We present both the standard coupling detector and the novel coupling detector. In Sec. \ref{sec:singlemodedetector} we consider a detector that couples to a single spatial momentum mode of the field, and its long interaction time limit.
In Sec.\ \ref{sec:conformaltime} we consider a modified detector whose dynamics is defined in terms of conformal time, rather than cosmological time. In all cases, we find that the detector's response exhibits a de-excitation peak, but the location and sharpness of this peak differ between the detector models. The detector whose dynamics is defined in terms of conformal time has a response closest to the Minkowski space response, as one might have expected. The detector whose dynamics is defined in terms of cosmological time shows peaking whose properties reflect both the expanding spatial volume and the field mode(s) to which the detector couples. 

In Sec.\ \ref{sec:noncompact} we discuss briefly the limit in which the spatial compactification is removed and the metric reduces to the spatially flat FLRW chart that covers half of de Sitter space. We find no de-excitation peaks, just a quadratic growth in the response at large de-excitation gaps, with the novel coupling giving a more rapid growth than the standard coupling. 

Finally, in Sec.\ \ref{sec:discrete}
we discuss the intuitive notion of the functionality of particle detectors. We introduce the unit-switching functions, which consist of a discrete sequence of intervals where the switching has constant magnitude, with instantaneous switch-on and switch-off moments for each interval. This allows us to directly link the Wightman function to the response function. Interestingly, the corresponding analysis shows, for the case of the novel coupling, that one is able to obtain a response function directly proportional to the number of on/off-switches if we choose suitable unit-switching functions with no mutual interference. However, the results for the standard coupling indicate that for the respective detector this is not possible, even if a reparametrization of the coupling time is considered again. 

Section \ref{sec:conclusions} gives a summary and the conclusions. Graphs displaying the numerical results are collected in Appendix~\ref{app:numerical}\null. Appendix \ref{app:dS-as} gives a technical proof of the de-excitation asymptotics described in Sec.~\ref{sec:noncompact}\null. 

We work in units in which $c = \hbar=1$, so that the squared line element has the physical dimension length squared, time has the physical dimension of length, and frequencies and energies have the physical dimension of inverse length. We use the mostly plus convention for the metric. 
In asymptotic formulae, 
$f(x)=O(x)$ denotes that $f(x)/x$ is bounded in the limit of interest, $f(x) = o(x)$ denotes that $f(x)/x\to0$ in the limit of interest, and $f(x) = o(1)$ denotes that $f(x)\to0$ in the limit of interest.

\section{Scalar field in compactified de Sitter spacetime}\label{sec:fieldandspacetime}

In this section, we consider a massive minimally coupled scalar field on a $(3+1)$-dimensional spacetime that is locally de Sitter, but is foliated by spatial sections that are flat and have the compact topology~$\mathbb{T}^3$. Adopting a Fock quantisation and selecting a vacuum, we shall take the opportunity to lay out the notation that will be used throughout the article.

\subsection{A scalar field in a spatially flat FLRW cosmology}

Let us start by considering a general spatially flat $\text{FLRW}$ metric, written as 
\begin{equation}\label{metric}
    ds^2= -dt^2 +a^2(t) \left(dx^2 +dy^2 +dz^2 \right),
\end{equation}
where the scale factor $a(t)$ is positive. We take $t$ and $a(t)$ to have the dimension of length and the spatial coordinates $x$, $y$ and $z$ to be dimensionless. Let $\eta$ be the conformal time that is related to the cosmological time via
\begin{equation}
    \frac{d\eta}{dt} = \frac{1}{a(t)}.
\label{eq:conftime-def}
\end{equation}
Note that $\eta$ is dimensionless. In the coordinates $(\eta,x,y,z)$, the metric reads 
\begin{equation}
    ds^2= a^2(\eta) \left( -d\eta^2 + dx^2 +dy^2 +dz^2 \right),
\end{equation}
denoting $a(t(\eta))$ by just $a(\eta)$ for brevity. 

The Lagrangian density of a massive minimally coupled scalar field $\phi$ is
\begin{equation}\label{Lagran}
    \mathscr{L} = -\frac{1}{2} \partial^{\mu} \phi \partial_{\mu} \phi - \frac{1}{2} m^2\phi^2,
\end{equation}
where $m\ge0$  denotes the mass of the field. The equation of motion (i.e. the Klein-Gordon equation) takes the form
\begin{equation}\label{phimov}
    \phi^{\prime \prime} - \hat{\mathbf{\Delta}} \phi + 2 \frac{a^{\prime}}{a} \phi^{\prime} + a^2m^2\phi = 0,
\end{equation}
where the prime denotes a derivative with respect to the conformal time $\eta$ and $\hat{\mathbf{\Delta}} = \partial_x^2 + \partial_y^2 + \partial_z^2$ is the Laplace-Beltrami operator of the conformal spatial metric $dx^2+dy^2+dz^2$. 
The Klein-Gordon inner product reads 
\begin{equation}\label{KGprod-phi}
\left(\phi_1, \phi_2\right) = i a^2 \int dx \, dy \, dz    
\left( \phi_1^{*} \partial_{\eta} \phi_2 - \phi_2 \partial_{\eta} \phi^{*}_1 \right) , 
\end{equation}
where the integral is evaluated on a hypersurface of constant~$\eta$ and the symbol $^{*}$ denotes the complex conjugate.

To solve the equations of motion, it is convenient to introduce the conformal field $\chi$ by 
\begin{equation}
    \chi(\eta, \vec{x}) = H a(\eta) \phi(\eta,\vec{x}),
\end{equation}
where $\vec{x} = (x,y,z)$, and the positive constant $H$ of dimension inverse length has been introduced for later convenience. 
Then, the equation of motion is simplified to 
\begin{equation} \label{ximov}
    \chi^{\prime \prime} - \hat{\mathbf{\Delta}} \chi + \mu^2(t) \, \chi = 0,
 \end{equation}
where $\mu^{2}(t) = ( a^2m^2
 - a^{\prime \prime}/a )$ is the
effective, time-dependent mass squared of the field~$\chi$.
In terms of the conformal field~$\chi$, the Klein-Gordon inner product reads 
\begin{equation}\label{KGprod-chi}
\left(\chi_1, \chi_2\right)_\chi = \frac{i}{H^2} \int dx \, dy \, dz    
\left( \chi_1^{*} \partial_{\eta} \chi_2 - \chi_2 \partial_{\eta} \chi^{*}_1 \right) . 
\end{equation}

\subsection{De Sitter scale factor and compactification of the spatial sections}\label{subsec:compactification}

Let us now specialise to a locally de Sitter spacetime by choosing 
\begin{equation}
a(t) = \frac{1}{H} e^{Ht} , 
\label{eq:a(t)}
\end{equation}
where $-\infty<t<\infty$, and the positive constant $H$ is the Hubble parameter, satisfying
\begin{equation}
    \frac{\dot{a}}{a} = H = \text{constant}. 
\end{equation}
Integrating Eq.~\eqref{eq:conftime-def}, 
we find 
\begin{equation}
\label{eq:eta-vs-t-and-a}
    a(\eta) = -\frac{1}{H\eta}, \quad \eta = - e^{-Ht}, 
\end{equation}
where $-\infty<\eta<0$: $\eta\to-\infty$ thus corresponds to the infinite past at $t\to-\infty$, and we have chosen the additive constant in $\eta$ so that $\eta=0$ corresponds to the future infinity at $t\to\infty$. 
The metric reads 
\begin{equation}
    ds^2= \frac{1}{H^2\eta^2}\left( -d\eta^2 + dx^2 + dy^2 + dz^2 \right).
\label{eq:conf-desittemetric}
\end{equation}
The effective mass squared of the conformal field $\chi$ takes the form
\begin{equation}
    \mu^2(\eta) = \left( \frac{m^2}{H^2}-2\right) \frac{1}{\eta^2}. 
\end{equation}
From now on we assume $m\ge \sqrt{2} H$, so that 
$\mu^2(\eta)\ge0$ for all~$\eta$. 

If $x$, $y$ and $z$ all have the fully infinite range, the spatially flat de Sitter metric \eqref{eq:conf-desittemetric} covers half of the de Sitter space, the hypersurfaces of constant $\eta$ have topology $\mathbb{R}^3$ and are Cauchy hypersurfaces for this half, 
and $\eta\to-\infty$ is a coordinate singularity on a null hypersurface~\cite{hawking-ellis}. We shall take each of $x$, $y$, and $z$ to be compactified on a circle, all three of them with the same period, so that the spacetime is given by Eq.~\eqref{eq:conf-desittemetric}
with the identification
\begin{equation}
    (\eta,x,y,z) \sim (\eta,x+L_0,y,z) \sim (\eta,x,y+L_0,z) \sim (\eta,x,y,z+L_0),
\label{eq:spat-identification}
\end{equation}
where $L_0$ is a dimensionless positive constant. 
The hypersurfaces of constant $\eta$ now have the compact topology of the three-torus~$\mathbb{T}^3$, and these hypersurfaces are Cauchy hypersurfaces for the spacetime, which has topology $\mathbb{R}\times \mathbb{T}^3$. 
The singularity at $\eta\to-\infty$ is no longer just a coordinate singularity but an orbifold-type singularity, related to but distinct from the singularity in Misner space~\cite{hawking-ellis}.

We emphasise that although the identification \eqref{eq:spat-identification} makes the spacetime \eqref{eq:conf-desittemetric} distinct from an open subset of de Sitter, the value of the positive dimensionless parameter $L_0$ in the identification has no coordinate-independent meaning. To see this, consider in Eq.~\eqref{eq:conf-desittemetric} the coordinate transformation 
\begin{equation}\label{isometry}
    (\tilde{\eta}, \vec{\tilde{x}}\,)
    =\left(\frac{L_1}{L_0} \eta ,\frac{L_1}{L_0} \vec{x} \right), 
\end{equation} 
where $L_1$ is another positive dimensionless constant: in the coordinates $(\tilde{\eta}, \vec{\tilde{x}}\,)$, the metric is as in Eq. \eqref{eq:conf-desittemetric} but with inserted overtildes, and $\tilde x$, $\tilde y$ and $\tilde z$ each have period~$L_1$. 
The underlying geometric reason for this is that translations in $x$, $y$ and $z$ in Eq.~\eqref{eq:conf-desittemetric} correspond to null rotation isometries in the embedding of $(3+1)$-dimensional de Sitter in $(4+1)$-dimensional Minkowski spacetime~\cite{hawking-ellis}, and null rotations do not have a Lorentz-invariant notion of a magnitude.
We could hence set (for example) $L_0=1$ without loss of generality. 
We shall, however, keep $L_0$ general in what follows, because $L_0$ provides a convenient way to encode the cosmological epoch in which a comoving detector operates, as we shall see in Secs.~\ref{sec:UDWdetector}, \ref{sec:singlemodedetector}
and~\ref{sec:conformaltime}. 
In particular, a detector operating in the late universe can be analysed as the $L_0\to\infty$ limit by techniques introduced in a related setting in Ref.~\cite{Continuum}, as we shall see in Sec.~\ref{sec:noncompact}. 

\subsection{Solution to the Klein-Gordon equation}

We solve the Klein-Gordon equation in terms of the conformal field $\chi$ by separation of variables. We seek a basis of positive Klein-Gordon norm mode functions with the ansatz  
\begin{equation}\label{solution}
\phi_{\vec{k}}(\eta,\vec{x}) = \frac{1}{H a(\eta)} \chi_{\vec{k}}(\eta,\vec{x}), 
\ \ \ 
    \chi_{\vec{k}}(\eta,\vec{x}) 
    = \frac{H}{L_0^{3/2}} \xi_{k}(\eta) \,e^{ i \vec{k}\cdot \vec{x}} , 
\end{equation}
where 
\begin{equation}\label{vec-k-def}
    \vec{k} = \frac{2\pi}{L_0} \vec{n} , 
\end{equation}
$\vec{n}=(n,l,r) \in \mathbb{Z}^3$ is an ordered triple of integers, $k = \sqrt{\vec{k}\cdot \vec{k}}$, and the dot indicates the scalar product in the conformal spatial metric $dx^2+dy^2+dz^2$. 
The time dependence is carried by the complex-valued functions~$\xi_{k}(\eta)$, 
for which the Klein-Gordon equation \eqref{ximov} reduces to
\begin{equation}\label{timedependence}
    \xi_{k}^{\prime\prime} + \left(\mu^{2}(\eta)+k^2 \right) \xi_{k} = 0 . 
\end{equation}
Note that as the coefficient function 
$\mu^{2}(\eta)+k^2$ in Eq.~\eqref{timedependence} depends on $\vec{k}$ only through the scalar wavenumber~$k$, it is consistent to assume the functions $\xi_{k}$ to be indexed by~$k$. 

We require the mode functions $\chi_{\vec{k}}$ to have positive Klein-Gordon inner product~\eqref{KGprod-chi}, where the integral over $x$, $y$ and $z$ now has the range $0\le x \le L_0$, $0\le y \le L_0$ and $0\le z \le L_0$, and to be orthonormal according to 
\begin{equation}\label{KGprod}
\left(\chi_{\vec{k}}, \chi_{\vec{k}'}\right)_\chi 
= \delta_{\vec{k} \vec{k}'},
\end{equation}
where $\delta_{\vec{k} \vec{k}'}$ denotes the Kronecker delta for the three labels $n$, $l$ and $r$, i.e. 
$\delta_{n n'}\delta_{l l'}\delta_{r r'}$.  
This is accomplished by requiring that 
$\xi_k$ satisfy the Wronskian condition 
\begin{equation} \label{relation}
   \xi_{k}^{*} \xi^{\prime}_{k} - \xi_{k}\xi_{k}^{\prime *} = -i . 
\end{equation}

The fields $\phi$ and $\chi$ can now be written as mode expansions in terms of the mode functions and their complex conjugates. For $\phi$, the expansion reads 
\begin{equation}
    \phi(\eta, \vec{x}) = \frac{1}{L_0^{3/2}} \frac{1}{a(\eta)} \sum_{\vec{k}} \left( b_{\vec{k}} \, \xi_{k}(\eta)\ e^{i \vec{k} \cdot \vec{x}} + b_{\vec{k}}^{*} \, \xi_{k}^{*}(\eta)\ e^{-i \vec{k} \cdot \vec{x}}\right),
\end{equation}
where $b_{\vec{k}}$ are complex-valued expansion coefficients and the sum over $\vec{k}$ denotes sums over the labels $n$, $l$ and~$r$.

\subsection{Fock quantisation of the field}

Let us promote in the standard manner the Fourier coefficients $b_{\vec{k}}$ and $b_{\vec{k}}^{*}$ to annihilation and creation operators, respectively, i.e. $\hat{b}_{\vec{k}}$ and $\hat{b}_{\vec{k}}^{\dagger}$, such that the Hilbert space of the quantized field is the usual Fock space. Then, the field operator takes the form
\begin{equation}
    \hat{\phi}(\eta, \vec{x}) = \frac{1}{L_0^{3/2}} \frac{1}{a(\eta)} \sum_{\vec{k}} \left( \hat{b}_{\vec{k}} \, \xi_{k}(\eta)\ e^{i \vec{k} \cdot \vec{x}} + \hat{b}_{\vec{k}}^{\dagger} \, \xi_{k}^{*}(\eta)\ e^{-i \vec{k} \cdot \vec{x}}\right),
\end{equation}
where the creation and annihilation operators satisfy the following canonical commutation relations

\begin{equation}
    \left[\hat{b}_{\vec{k}}, \, \hat{b}_{\vec{k}'}^{\dagger}  \right] = \delta_{\vec{k} \vec{k}'} \hat{\mathbb{I}}, \quad \left[\hat{b}_{\vec{k}}, \, \hat{b}_{\vec{k}'} \right] = \left[\hat{b}_{\vec{k}}^{\dagger}, \, \hat{b}_{\vec{k}'}^{\dagger}  \right] = 0, 
\end{equation}
and we have used $\hat{\mathbb{I}}$ to denote the identity operator. 

\subsection{Choice of vacuum}

Noticing Eq.~\eqref{timedependence}, 
let us define
the time-dependent effective frequency $\omega_{k}(\eta)$ by
\begin{equation}
    \omega_{k}(\eta) = \sqrt{\mu^2(\eta) 
    + k^2 }.
\end{equation}
For the modes with $k>0$, we have $\omega_{k}(\eta) \to k>0$ as $\eta\to-\infty$, and a distinguished early time vacuum can be identified by choosing $\xi_{k}(\eta)$ to satisfy the adiabatic criterion 
\begin{equation}\label{eq:adiabatic-form}
    \xi_{k}(\eta) \xrightarrow[]{\text{as}\;\eta \rightarrow -\infty } \frac{1}{\sqrt{2 \ \omega_{k}(\eta)}} e^{-i\int^{\eta}d\bar{\eta} \  \omega_{k}(\bar{\eta})}, 
\end{equation}
which satisfies Eq.~\eqref{relation}, and has at $\eta\to-\infty$ the leading behaviour $\xi_{k}(\eta) \sim  e^{ik|\eta|}/\sqrt{2k}$, reducing to Minkowski-like oscillations in the conformal time. This state is known as the Bunch-Davies state~\cite{Mukhanov,Bunch:1978yq}. 
The exact solution to Eq.~\eqref{timedependence} with this asymptotic form is
\begin{equation}\label{Hankel1}
    \xi_{k}(\eta) = \sqrt{\frac{|\eta|\pi}{4}} H^{(1)}_{\nu} ( k |\eta|), \quad \text{with} \quad \nu = \sqrt{\frac{9}{4}-\frac{m^2}{H^2}},
\end{equation}
where $H^{(1)}_{\nu}$ is the Hankel function of the first kind~\cite{dlmf}. 

The mode with $k=0$, i.e.\ $n=l=r=0$, is qualitatively different, since $\omega_{0}(\eta) \to 0$ as $\eta\to-\infty$, and the adiabatic criterion \eqref{eq:adiabatic-form} cannot be imposed. This mode is known as a massive zero mode, and it needs a special treatment~\cite{ZeroMode,TovLou}. The massive zero mode is, however, not of interest for the present paper, as discussed in the Introduction, and from now on we will just omit it from the field. 

For concreteness, we make from now on a further simplification by setting $m=\sqrt{2}\,H$, which means that the field equation is identical to that of a massless conformally coupled field. $\xi_{k}(\eta)$, given in Eq.~\eqref{Hankel1} is then expressible in terms of elementary functions as 
\begin{equation} \label{unitarydeSitterEvol}
    \xi_{k}(\eta) = -i\sqrt{\frac{1}{2 k}} e^{i k |\eta|}.
\end{equation}

\subsection{Wightman function}

We can write the Wightman function of the field $\phi$ as
\begin{equation}
    W^{\phi}(\eta,\vec{x}; \eta', \vec{x}') = W^{\phi}_0(\eta, \eta') + W^{\phi}_{\modes}(\eta,\vec{x}; \eta',\vec{x}'),
\end{equation}
where $W^{\phi}_0(\eta, \eta')$ denotes the Wightman function for the zero mode, and $W^{\phi}_{\modes}(\eta,\vec{x}; \eta',\vec{x}')$ is the Wightman function of the rest of the modes, on which we focus our attention. Recalling that $W^{\phi}(\eta,\vec{x}; \eta',\vec{x}') = \bra{0} \hat{\phi}(\eta,\vec{x}), \hat{\phi} (\eta',\vec{x}') \ket{0}$, we find
\begin{equation}
    W^{\phi}_{\modes}(\eta,\vec{x}; \eta',\vec{x}') = \frac{1}{L_0^{3}a(\eta)a(\eta')}
\sideset{}{'}\sum_{\vec{k}} \xi_{k}(\eta)\xi^{*}_{k}(\eta')  e^{i\vec{k}\cdot(\vec{x}-\vec{x}')}, 
\end{equation}
where the prime on the sum denotes that $\vec{k} = \vec{0}$ is omitted from the sum. 
Substituting the expressions for $a(\eta)$, $\xi_{k}(\eta)$ and~$\vec{k}$, we obtain
\begin{equation}\label{ConvWeight}
W^{\phi}_{\modes}(\eta,\vec{x}; \eta',\vec{x}') =\frac{H^2 \ \eta \eta'}{2 L_0^{3}} \sideset{}{'}\sum_{\vec{k}} \frac{1}{k} e^{i k \left(|\eta|- |\eta'|\right)}
    \, e^{ i \vec{k} \cdot (\vec{x}-\vec{x}')} .
\end{equation}

As a consistency check, we note that the expression \eqref{ConvWeight} is invariant under the coordinate transformation \eqref{isometry} and the concomitant rescaling $L_0\to L_1$, recalling that 
$k L_0 = 2\pi \sqrt{n^2+l^2+r^2} = \tilde{k}  L_1$ under this transformation. In what follows, we shall use this invariance to encode the time dependence of the detector's response in~$L_0$. 

\section{Unruh-DeWitt point-like detector} \label{sec:UDWdetector}

In this section, we investigate an Unruh-DeWitt detector coupled to the massive scalar field~\cite{Unruh:1976db,DeWitt:1980hx}. This detector is a spatially point-like two-level system, moving on the prescribed timelike worldline $\mathrm{x}(\mathscr{T}) = \left( t(\mathscr{T}),\vec{x}(\mathscr{T})\right)$, with $\mathscr{T}$ being 
the detector's 
proper time, and $\vec{x}$ referring to the three-dimensional spatial coordinates. The detector's Hilbert space $\mathcal{H}_D$ is spanned by the orthonormal states $\left\lbrace \ket{0}_D, \ket{E}_D\right\rbrace$, which satisfy $\hat{\mathrm{H}}_D \ket{0}_D = 0$ and $\hat{\mathrm{H}}_D \ket{E}_D = E\ket{E}_D$, where $\hat{\mathrm{H}}_D$ is the detector's Hamiltonian operator and $E$ is a real-valued constant. For $E>0$, the state $\ket{0}_D$ is the detector's ground state and $\ket{E}_D$ is the excited state; for $E<0$, the roles of $\ket{0}_D$ and $\ket{E}_D$ are reversed. We may refer to $E$ as the detector's energy gap. In the special case $E=0$, both states have zero energy.

Let $\mathcal{H}_{\phi}$ be the Fock space of the scalar field~$\phi$. The Hilbert space of the detector-field system is $\mathcal{H}_D \otimes \mathcal{H}_{\phi}$. The Hamiltonian operator of the total system is $\hat{\mathrm{H}}= \hat{\mathrm{H}}_D + \hat{\mathrm{H}}_{\phi} + \hat{\mathrm{H}}_{\text{int}}$, where $\hat{\mathrm{H}}_{\phi}$ is the Hamiltonian arising from Eq. \eqref{Lagran} by a Legendre transform and $\hat{\mathrm{H}}_{\text{int}}$ denotes the interaction Hamiltonian operator between the detector and the field. 

\subsection{Standard coupling}
\label{subsec:standardcoupling-response}

We begin our study by reproducing the well-known results for a coupling of the form
\begin{equation} \label{Conventional}
\hat{\mathrm{H}}_{\text{int}} = \epsilon \,\Theta(\PT) \hat{\mu}(\PT)\otimes
\sideset{}{'}\sum_{\vec{k}}
\hat{\phi}_{\vec{k}}(\mathrm{x}(\mathscr{T})),
\end{equation}
where we recall that the prime on the sum denotes that $\vec{k}=\vec{0}$ is not included, and where $\epsilon$ is a real-valued coupling constant, assumed to be small, such that the Dyson expansion of the time-evolution operator of the detector-field system states can be identified as a valid approximation to the first order. $\Theta(\mathscr{T})$ denotes a real-valued switching function which specifies when the detector (and thus the interaction) is turned on and off, and $\hat{\mu}(\mathscr{T})$ is the detector (field) operator, which is considered to be a monopole moment such that it interacts at one particular point at a time along the worldline trajectory. Hence, the evolution of this monopole moment operator in the interaction picture is given by
\begin{equation}
    \hat{\mu}(\mathscr{T}) = e^{i\mathscr{T}\hat{\mathrm{H}}_D} \,\hat{\mu}(\mathscr{T}_0) \,e^{-i\mathscr{T}\hat{\mathrm{H}}_D}.
\end{equation}

We assume that before the interaction begins, the detector is prepared in the state $\ket{0}_D$ and the field is prepared in the Bunch-Davies vacuum described above. In first-order perturbation theory, the probability to find the detector in the state $\ket{E}_D$ after the interaction has ceased, regardless of the final state of the field, is \cite{BD,Crispino,Wald,Junker:2001gx}
\begin{equation}
    \mathscr{P}(E) = \epsilon^2 \,\big| {}_D \! \bra{E}\hat{\mu}(\PT_0) \ket{0}_D \big|^2 \, \mathcal{F}^{\phi}_{\modes}(E),
\end{equation}
where $\mathcal{F}^{\phi}_{\modes}(E)$ is the detector's response function, given by 
\begin{equation}\label{genresp}
    \mathcal{F}^{\phi}_{\modes}(E) =  \int d\PT \, d\PT' \, \Theta(\PT)\Theta(\PT') e^{-iE(\PT-\PT')} W_{\modes}(\PT,\PT'),
\end{equation} 
and $W_{\modes}(\PT,\PT')$ denotes the pullback of the field's Wightman function in the Bunch-Davies state to the detector's trajectory~$\mathrm{x}(\PT)$, with the $\vec{k}=\vec{0}$ contribution omitted.

We shall consider detector's trajectories that are comoving with the cosmological expansion, i.e.\ $\mathrm{x}(\PT)= (\PT , \vec{0})$, where the detector's proper time $\PT$ is equal to the cosmological time~$t$, and the spatial position can be chosen to be at $\vec{x} = \vec{0}$ without loss of generality by the spatial homogeneity of the spacetime and of the Bunch-Davies state.
We choose the switching function to be 
\begin{equation} \label{switching}
    \Theta(t) = \left\{
     \begin{array}{@{}l@{\thinspace}l}
       \cos^4 \! \left( \frac{\pi \left(t-t_{\text{mid}}\right) }{2\Delta}\right)  &\text{ if } \ |t-t_{\text{mid}}| \leq \Delta ,\\
       0 &  \text{ otherwise},
     \end{array}
   \right.
\end{equation}
where the positive parameter $\Delta$ is half the interaction duration, and the parameter $t_{\text{mid}}$ denotes the cosmological time at the midpoint of the interaction interval. This switching function is a close approximation to a Gaussian over its duration~\cite{Cong:2020crf}, and it is smooth everywhere except at the switch-on and switch-off moments, where it is~$C^3$. 

The response function $\mathcal{F}^{\phi}_{\modes}(E)$ is now obtained by substituting the oscillator mode Wightman function \eqref{ConvWeight}, the detector's trajectory, and the switching function \eqref{switching} into the response function formula~\eqref{genresp}. 
$\mathcal{F}^{\phi}_{\modes}(E)$ depends a priori on five parameters. These are 
the Hubble parameter $H$ and the detector's energy gap~$E$, which have dimension inverse length, the interaction duration parameter $\Delta$ and the interaction interval midpoint~$t_{\text{mid}}$, which have dimension length, and the spatial compactification parameter~$L_0$, which is dimensionless. 
$\mathcal{F}^{\phi}_{\modes}(E)$ itself is dimensionless, as a consequence of the switching function \eqref{switching} being dimensionless. 
By dimensional analysis, it follows that $\mathcal{F}^{\phi}_{\modes}(E)$ can depend on the five parameters only through four dimensionless combinations, for which we can choose~$E/H$, which is the energy gap in units of the Hubble parameter, $H\Delta$, which is half the interaction duration in units of the inverse Hubble parameter, 
$Ht_{\text{mid}}$, 
which is the cosmological time at the interaction interval midpoint in units of the inverse Hubble parameter, and the spatial compactification parameter~$L_0$. 
This is however still an overcounting, because $L_0$ does not have a coordinate-invariant meaning, as discussed in Sec.~\ref{subsec:compactification}: 
the dependence on $Ht_{\text{mid}}$ and $L_0$ enters only through the combination 
$L_0 e^{Ht_{\text{mid}}}$, which is the cube root of the spatial volume at the interaction interval midpoint, and is thus a coordinate-invariant quantity. We may therefore remove the redundancy in the parametrisation by setting $t_{\text{mid}}=0$, which gives $L_0$ a geometric meaning as the cube root of the spatial volume at the interaction interval midpoint. In particular, large values of $L_0$ then mean that the detector operates in the late universe, and small values of $L_0$ mean that the detector operates in the early universe. 

In this notation, we find 
\begin{align} 
    \mathcal{F}^{\phi}_{\modes}(E) &= \frac{1}{2L_0^3}\sideset{}{'}\sum_{\vec{k}} \frac{1}{k} \, \Biggl| \int_{-\Delta}^{\Delta} dt \, 
    H \cos^{4} \! \left( \frac{\pi t }{2 \Delta}\right){e^{-t(H+iE)}} 
    e^{ i k \exp( -Ht)}\Biggr|^2
    \notag\\
    &= \frac{1}{2L_0^3}\sideset{}{'}\sum_{\vec{k}} \frac{1}{k} \, \Biggl| \int_{-H\Delta}^{H\Delta} d\tau \, \cos^{4} \! \left( \frac{\pi \tau }{2 H \Delta}\right)  
    e^{-\tau(1+i E/H)} e^{ i k \exp( -\tau )}\Biggr|^2 ,
    \label{resp}
\end{align}
where the last equality comes by 
introducing the new dimensionless proper time integration variable $\tau = Ht$.
From the last expression in Eq.~\eqref{resp} it is plain that the independent parameters in $\mathcal{F}^{\phi}_{\modes}(E)$ are $E/H$, $H\Delta$ and $L_0$, 
of which $L_0$ enters both via the overall factor $L_0^{-3}$ and via the dependence of $\vec{k}$ on~$L_0$ given by Eq.~\eqref{vec-k-def}. 

We note that the sum over $\vec{k}$ in Eq.~\eqref{resp} converges because the summand is bounded at large $k$ by a constant times~$k^{-11}$: as the exponent in this falloff is more negative than~$-3$, convergence follows by comparison with the three-dimensional integral over~$\vec{k}$. To verify the large $k$ falloff of the summand, consider the integral over $\tau$ inside the modulus squared in Eq.~\eqref{resp}, and integrate by parts repeatedly, integrating the factor 
$e^{-\tau} e^{ i k \exp( -\tau )}$ \cite{Wong:1989}. 
In the first four integrations by parts, the substitution terms vanish, because the $\cos^4$ factor and its first three derivatives vanish at the upper and lower limits. The fifth integration by parts creates substitution terms proportional to $k^{-5}$, and the remaining integral has a large $k$ asymptotic expansion that proceeds in higher negative integer powers of $k$ \cite{Wong:1989}. Combining, the summand is at large $k$ bounded by a constant times 
$k^{-1} {(k^{-5})}^2 = k^{-11}$.

\subsection{Novel coupling}
\label{subsec:novelcoupling-response}

Let us now introduce our alternative proposal for the detector coupling in an expanding cosmology. As discussed in the Introduction, this novel detector is intended to take into account a contribution of the background evolution to the standard field dynamics, such that the particles that are being probed correspond to the ones with \textit{proper} dynamics. In other words, we aim to design a detector that can absorb excitations due to the background dynamics, addressing the corresponding readings to particles with dynamics sourced from the field's \textit{own} evolution. Thus, we can leverage a subtle freedom in the design of particle detectors and introduce a time-dependent scaling to the coupling. Moreover, in order to consider the setup as feasible as possible, we shall require, as stated in the Introduction, that the information recompiled by the detector comes uniquely from measuring spacetime quantities, such as metric functions. 

Motivated by the massless case, where the conformal field $\chi$ action is equivalent to that of a field in flat spacetime decoupled from dynamical contributions of the geometry, we could think of the massive conformal field dynamics as being \textit{minimally} or \textit{weakly} influenced by the background dynamics, since the kinetic term of its Lagrangian (from which the differential part of the equations of motion is derived) is left unaffected, i.e. remains as in the Minkowski case\footnote{In fact, this motivation is reinforced by Ref.\ \cite{Triada} where the conformal field dynamics is discussed and proven to be unitarily implementable. Sticking to that definition of unitarity, we would be able to isolate in the detection the unitarily implementable part of the time evolution of the scalar field for a spacetime with constant but nonvanishing four-dimensional curvature.}. Therefore, we shall consider for the new detector a coupling to the conformal field instead, which mathematically is equivalent to choosing the scale factor $a(t)$ as the time-dependent scaling.

For the above reasons, we consider the novel interaction Hamiltonian $\hat{\mathrm{H}}_{\text{int}}^{\chi}$, 
given by rescaling the conventional interaction Hamiltonian $\hat{\mathrm{H}}_{\text{int}}$ \eqref{Conventional} by 
\begin{equation} \label{Interact}
    \hat{\mathrm{H}}_{\text{int}}^{\chi} 
    =
    (L_0 H a)
    \hat{\mathrm{H}}_{\text{int}}
    = 
    \epsilon \, L_0\,\Theta(\PT) \hat{\mu}(\PT)\otimes
    \sideset{}{'}\sum_{\vec{k}}
    \hat{\chi}_{\vec{k}}(\mathrm{x}(\mathscr{T})) , 
\end{equation}
where the constant $L_0$ has been included in order to maintain invariance under the coordinate transformation~\eqref{isometry}. 
Using Eq.~\eqref{eq:eta-vs-t-and-a}, 
and following the steps in Sec.~\ref{subsec:standardcoupling-response}, 
we find that the response function for a detector on a comoving trajectory is given by 
\begin{equation}\label{chiresponse}
   \mathcal{F}^{\chi}_{\modes}(E)  = \frac{1}{2 L_0} \sideset{}{'}\sum_{\vec{k}} \frac{1}{k} \, \Biggl| \int_{-H\Delta}^{H\Delta} d\tau \,  \cos^{4} \! \left(  \frac{\pi \tau }{2 H \Delta}\right) 
   e^{-i\tau E/H} e^{ i k \exp(-\tau)} 
   \Biggr|^2 . 
\end{equation}
As with the conventional coupling response~\eqref{resp}, it is plain from Eq.~\eqref{chiresponse} that the independent parameters in $\mathcal{F}^{\chi}_{\modes}(E)$ are $E/H$, $H\Delta$ and $L_0$. 
$L_0$~again has a geometric meaning as the cube root of the spatial volume at the interaction interval
midpoint. 
Convergence of the sum over $\vec{k}$ in Eq.~\eqref{chiresponse} can be verified as for the sum in Eq.~\eqref{resp}.

\subsection{Numerical results}
\label{subsec:conv+novel:numerical}

Figures \ref{fig:ConvA}--\ref{fig:EvolB} in Appendix \ref{app:numerical} show numerical plots 
for $\mathcal{F}^{\phi}_{\modes}(E)$
and 
$\mathcal{F}^{\chi}_{\modes}(E)$
from Eqs.~\eqref{resp} and~\eqref{chiresponse}. 

First, we note that while the sum over $\vec{k}$ is convergent, a large $k$ cut-off needs to be introduced in the numerical evaluation. Figures \ref{fig:ConvA} and \ref{fig:ConvB} give evidence that within the parameter range used in the plots, a low cutoff value for $|n|$, $|l|$ and $|r|$ suffices for visually accurate figures. 
Figures 
\ref{fig:EvolA} and \ref{fig:EvolB}
show plots of 
$\mathcal{F}^{\phi}_{\modes}(E)$
and 
$\mathcal{F}^{\chi}_{\modes}(E)$
as a function of the dimensionless gap $E/H$ for fixed~$H\Delta = 3.5$, for selected values of~$L_0$, with the cutoff values stated in the captions. The plots show clearly that excitations are suppressed compared with de-excitations, which was to be expected. 
They also indicate that the de-excitation probability has a peak at a characteristic $L_0$-dependent negative value of $E/H$, and as $L_0$ increases, the location of the peak approaches zero and the peaking becomes sharper. Recalling the interpretation of $L_0$ in terms of the age of the universe, this means that the de-excitation peaking is sharper and closer to zero gap for a detector operating in the late universe. Comparing 
Figs.~\ref{fig:EvolA} and~\ref{fig:EvolB}, 
we see that the novel coupling detector exhibits a stronger peaking than the standard coupling detector.

\section{Single-mode detector}
\label{sec:singlemodedetector}

In this section, we consider versions of the standard-coupling and novel-coupling detectors that couple to exactly one of the modes of the field, with the finite-duration switching \eqref{switching} and with its infinite-duration limit $\Delta\to\infty$.

\subsection{Finite interaction duration}
\label{subsec:singlemode-finiteduration}

Consider a detector defined as above but assumed to couple to exactly one spatial mode of the field, with $\vec{k} = \vec{k}_0 = 2\pi (n_0,l_0,r_0)/L_0$. 
Truncating the sums in 
Eqs.~\eqref{resp}
and
\eqref{chiresponse}
to one mode, 
we see that the single-mode response functions are given by 
\begin{subequations}
\label{eq:singlemode-responses}
\begin{align} 
    \mathcal{F}^{\phi}_{k_0}(E) &= \frac{1}{2L_0^3 k_0} \, \Biggl| \int_{-H\Delta}^{H\Delta} d\tau \, \cos^{4} \! \left( \frac{\pi \tau }{2 H \Delta}\right)  
    e^{-\tau(1+i E/H)} e^{ i k_0 \exp( -\tau )}\Biggr|^2, 
    \label{resp-singlemode}
    \\
   \mathcal{F}^{\chi}_{k_0}(E) &= \frac{1}{2 L_0 k_0}  \, \Biggl| \int_{-H\Delta}^{H\Delta} d\tau  \,  \cos^{4} \! \left(  \frac{\pi \tau }{2 H \Delta}\right) 
   e^{-i\tau E/H} e^{ i k_0 \exp(-\tau)} 
   \Biggr|^2 , 
   \label{chiresponse-singlemode}
\end{align}
\end{subequations}
where $k_0= |\vec{k}_0| = 2\pi \sqrt{n_0^2+l_0^2+r_0^2}/L_0$. 

Figures 
\ref{fig:OneModeA} and \ref{fig:OneModeB} 
show plots of 
$\mathcal{F}^{\phi}_{k_0}(E)$
and 
$\mathcal{F}^{\chi}_{k_0}(E)$ 
as a function of the dimensionless gap $E/H$ for the lowest momentum mode, 
$k_0 = 2\pi/L_0$, 
for fixed $H\Delta = 3.5$, 
with $L_0=0.5$ and $L_0=1.5$. 
The qualitative features are similar to those seen for $\mathcal{F}^{\phi}_{\modes}(E)$
and 
$\mathcal{F}^{\chi}_{\modes}(E)$ in Figs.~ \ref{fig:EvolA} and~\ref{fig:EvolB}, including the de-excitation peak at negative~$E/H$, and the dependence of the locus and shape of this peak on~$L_0$. 
For 
$\mathcal{F}^{\phi}_{k_0}(E)$
and~$\mathcal{F}^{\chi}_{k_0}(E)$, we can however gain more analytic understanding of these features, as we shall now discuss. 

First, $\mathcal{F}^{\phi}_{k_0}(E)$
and $\mathcal{F}^{\chi}_{k_0}(E)$ 
both have the falloff $\mathcal{O} \! \left( E^{-10} \right)$ as $E\to\pm\infty$. This follows by repeated integration by parts in the integrals over $\tau$ in Eq.~\eqref{eq:singlemode-responses}, integrating the factor $e^{-i\tau E/H}$, and noting that the $\cos^4$ factor and its first three derivatives vanish at the endpoints of the integration~\cite{Wong:1989}. 

Second, consider the location of the de-excitation peak. In the integrals over $\tau$ in Eq.~\eqref{eq:singlemode-responses}, 
the phase of the integrand is $-(E/H)\tau + k_0 e^{-\tau}$, and for $|\tau| \lesssim 1$ this can be approximated by $-\bigl((E/H) + k_0\bigr)\tau$ up to an additive constant. This suggests that when $H\Delta$ is not much larger than order unity, the integral is largest when the phase is approximately constant over $|\tau| \lesssim 1$, that is, for $E/H \approx - k_0$. 
For the values used in Figs.~\ref{fig:OneModeA} and~\ref{fig:OneModeB}, this gives the respective estimates $E/H \approx -12$ for Fig.~\ref{fig:OneModeA} and $E/H \approx -4$ for Fig.~\ref{fig:OneModeB}. The plots show that this estimate is fairly accurate for $\mathcal{F}^{\chi}_{k_0}(E)$ with the parameters in Fig.~\ref{fig:OneModeB}, and less accurate but correct within the order of magnitude in the other cases.

\subsection{Infinite interaction duration}
\label{subsec:singlemode-infiniteduration}

We now consider the single-mode response functions 
$\mathcal{F}^{\phi}_{k_0}(E)$
and $\mathcal{F}^{\chi}_{k_0}(E)$ given by Eq.~\eqref{eq:singlemode-responses}
in the limit of long interaction duration, $\Delta\to\infty$. 

Consider first the novel coupling response function $\mathcal{F}^{\chi}_{k_0}(E)$~\eqref{chiresponse-singlemode}. In the $\Delta\to\infty$ limit, the integral over $\tau$ under the modulus squared becomes 
\begin{align}
& \ \ \ \ 
\int_{-\infty}^{\infty} d\tau \ e^{-i(E/H)\tau} \, \exp \! \left( i k_0 e^{-\tau}  \right) 
\notag\\
& =
\int_0^{\infty} \frac{d\rho}{\rho} \ \rho^{iE/H} \, e^{i k_0 \rho }
\notag\\
& =
e^{-\frac12 \pi E/H}  \, e^{-i{(E/H)}\ln k_0}  \,  
\int_0^{\infty} \frac{d\sigma}{\sigma} \sigma^{iE/H} e^{-\sigma}
\notag\\
& =
e^{-\frac12 \pi E/H}  \, e^{-i{(E/H)}\ln k_0}  \,  
\Gamma \! \left( \frac{iE}{H}\right), 
\label{eq:nonstand-infinite-durationi-integral}
\end{align}
where the first equality comes by the substitution $\rho= e^{-\tau}$, 
the second equality comes by deforming the integration contour to $\rho = i \sigma/k_0$, and the third equality follows by the integral representation of Euler's gamma-function $\Gamma$~\cite{dlmf}. 
Taking the modulus squared and using the identity 
${|\Gamma(iy)|}^2 = \pi/[y \sinh( \pi y)]$, 
valid for $y\in\mathbb{R}\setminus\{0\}$~\cite{dlmf}, we find 
\begin{equation}\label{eq:novelcoupling-longtime}
\mathcal{F}^{\chi}_{k_0,\infty}(E) = 
\lim_{\Delta \to \infty} \mathcal{F}^{\chi}_{k_0}(E)
= \frac{\pi}{ k_0 L_0} \frac{H}{E}\frac{1}{e^{2\pi E/H}-1}.
\end{equation}

For the conventional coupling response function $\mathcal{F}^{\phi}_{k_0}(E)$~\eqref{resp-singlemode}, proceeding similarly and using the identity $\Gamma(x+1) = x \, \Gamma(x)$ \cite{dlmf} gives 
\begin{equation}\label{eq:standardcoupling-longtime}
\mathcal{F}^{\phi}_{k_0,\infty}(E)  =
\lim_{\Delta \to \infty} \mathcal{F}^{\phi}_{k_0}(E)
= \frac{\pi}{ k_0 L_0^3}\frac{E}{H} \frac{1}{e^{2\pi E/H}-1}. 
\end{equation}

Plots of $\mathcal{F}^{\chi}_{k_0,\infty}(E)$ and $\mathcal{F}^{\phi}_{k_0,\infty}(E)$ are shown in Fig.~\ref{fig:Delta}. 
$\mathcal{F}^{\chi}_{k_0,\infty}(E)$ and $\mathcal{F}^{\phi}_{k_0,\infty}(E)$ 
are both thermal, 
in the temperature 
\begin{align}
T_{\text{dS}} = \frac{H}{2\pi} , 
\label{eq:GH-temp}
\end{align}
in the sense that they satisfy the detailed balance condition \cite{Einstein-detailedbalance,terhaar-book} 
\begin{align}
\mathcal{F}(- E) 
= e^{E/T_{\text{dS}}} \mathcal{F}(E) . 
\end{align}
Here, $T_{\text{dS}}$ \eqref{eq:GH-temp} is the Gibbons-Hawking temperature that characterises inertial observers' response to the Bunch-Davies state in full de Sitter spacetime 
\cite{Bunch:1978yq,Gibbons:1977mu}. 
$\mathcal{F}^{\chi}_{k_0,\infty}(E)$ and $\mathcal{F}^{\phi}_{k_0,\infty}(E)$ however differ in their detailed functional form. 
$\mathcal{F}^{\phi}_{k_0,\infty}(E)$ has the Planckian profile that arises for an inertial detector coupled to a full massless field in the Bunch-Davies vacuum in full four-dimensional de Sitter spacetime \cite{Gibbons:1977mu}, 
and also in the 
Unruh effect in four-dimensional Minkowski spacetime 
\cite{Unruh:1976db,Takagi}, whereas $\mathcal{F}^{\chi}_{k_0,\infty}(E)$ has a Planckian profile that arises in the Unruh effect in two spacetime dimensions~\cite{Takagi}. 
The differences between the two show up especially at large negative~$E$, where $\mathcal{F}^{\phi}_{k_0,\infty}(E)$ diverges linearly in $E$ while $\mathcal{F}^{\chi}_{k_0,\infty}(E)$ falls off as~$1/E$, 
and near $E=0$, where $\mathcal{F}^{\phi}_{k_0,\infty}(E)$ 
remains regular but $\mathcal{F}^{\chi}_{k_0,\infty}(E)$ diverges as~$1/E^2$. 

Note also that when $\vec{k_0} = 2\pi(n_0,l_0,r_0)/L_0$ and the triple $(n_0,l_0,r_0)$ is fixed, $\mathcal{F}^{\chi}_{k_0,\infty}(E)$ is independent of~$L_0$, which means that the response is the same in the early universe and in the late universe, whereas $\mathcal{F}^{\phi}_{k_0,\infty}(E)$ is proportional to~$1/L_0^2$, meaning that the response has a smaller overall magnitude in the late universe.

\section{Detector with conformal time microphysics} \label{sec:conformaltime}

The results obtained above for the standard and novel couplings in Eqs.~\eqref{Conventional} and \eqref{Interact} show that the time dependence plays a fundamental role in the two cases, even for the comoving detectors that we are considering. One may then wonder whether the results depend drastically on the time chosen to describe the detector evolution. In an effort to understand different characteristics of the detectors, we shall now analyse a different parametrisation of the time. We shall study the conformal time, a natural alternative that can be found in the literature, for example in Refs.~\cite{Ralph, Olson, Quach:2021vzo}.

\subsection{Conformal time microphysics} 

We take the static Hamiltonian $\hat{\mathrm{H}}_D$ to determine the evolution in conformal time~$\eta$, i.e.
\begin{equation}\label{confo1}
    \hat{\mathrm{H}}_D \ket{\Psi}_D = i \frac{d}{d\eta} \ket{\Psi}_D = i a(t) \frac{d}{dt} \ket{\Psi}_D . 
\end{equation}
This amounts to rescaling the static Hamiltonian by the inverse of the scale factor. The evolution of the monopole moment operator in conformal time then reads 
\begin{equation}
    \hat{\mu}(\eta) = e^{i\hat{\mathrm{H}}_D (\eta-\eta_0)} \hat{\mu}(\eta_0) \, e^{-i\hat{\mathrm{H}}_D (\eta-\eta_0)}.
\end{equation}
We take the total system to be initially prepared in the state $\ket{0}_D\otimes \ket{0}_{\phi}$, and we adapt the analysis of Sec.\ \ref{sec:UDWdetector} to the evolution in the conformal time. We consider both the conventional coupling, similar to that in Eq.~\eqref{Conventional}, and the novel coupling similar to that in Eq.~\eqref{Interact}.
After the interaction has ceased, the detector's probability to have made a transition is a multiple of the response function, which is now given by 
\begin{subequations}
\begin{align}
\mathcal{F}^{\phi}(E) 
& = 
\int d\eta \, d\eta' \, \Theta(\eta)\Theta(\eta') \, e^{-iE(\eta-\eta')} \, W(\eta,\eta'),
\label{eq:conftime-F-phi}
\\
\mathcal{F}^{\chi}(E) 
&= 
L_0^2 \int d\eta \, d\eta' \, \Theta(\eta)\Theta(\eta') \, e^{-iE(\eta-\eta')} \, {(\eta \eta')}^{-1}W(\eta,\eta'),
\label{eq:conftime-F-chi}
\end{align}
\end{subequations}
where the superscript $\phi$ refers to the conventional coupling and the superscript $\chi$ refers to the novel coupling. $W(\eta,\eta')$ is the pullback of the field's Wightman function to the detector's comoving trajectory, and the switching function $\Theta$ is now a function of the conformal time.

For the switching function, we take 
\begin{equation} \label{switchingConf}
    \Theta(\eta) = \left\{
     \begin{array}{@{}l@{\thinspace}l}
       \cos^4 \! \left( \frac{\pi (\eta-\eta_{\text{mid}}) }{2\Delta}\right)  &\text{ if } \ |\eta-\eta_{\text{mid}}| \leq \Delta ,\\
       0 &  \text{ otherwise}.
     \end{array}
   \right.
\end{equation}
To compare with the results of Sec.~\ref{sec:singlemodedetector}, we consider a coupling to exactly one spatial mode of the field, with $\vec{k} = \vec{k}_0 = 2\pi (n_0,l_0,r_0)/L_0 \ne (0,0,0)$. 
We write $k_0= |\vec{k}_0| = 2\pi \sqrt{n_0^2+l_0^2+r_0^2}/L_0 >0$. 

\subsection{Novel coupling}\label{SecVB}

For the novel coupling, substituting the single-mode contribution to Eq.~\eqref{ConvWeight} in Eq.~\eqref{eq:conftime-F-chi} gives 
\begin{align}
\mathcal{F}^{\chi}_{k_0}(E) 
&= 
\frac{H^2}{2 k_0 L_0} \Bigg| \int_{\eta_{\text{mid}}-\Delta}^{\eta_{\text{mid}}+\Delta} d\eta \, \cos^4 \! \left( \frac{\pi (\eta-\eta_{\text{mid}}) }{2\Delta}\right) e^{-i\left( E+  k_0\right) \eta} \Bigg|^2
\notag
\\
&= 
\frac{H^2}{2 k_0 L_0} \Bigg| \int_{-\Delta}^{\Delta} d\sigma \, \cos^4 \! \left( \frac{\pi \sigma}{2\Delta}\right) e^{-i\left( E+  k_0\right) \sigma} \Bigg|^2 , 
\label{eq:conf-chi-F}
\end{align}
where the last expression has come by the change of variables $\eta = \sigma + \eta_{\text{mid}}$. 
Note that $\eta_{\text{mid}}$ has dropped out of Eq.~\eqref{eq:conf-chi-F}, and while $k_0$ and $L_0$ only appear in the combination $k_0 L_0$ in the prefactor, $k_0$ appears alone in the exponent. This means that the detector model invokes knowledge beyond the metric in the sense that it knows about the value of the compactification parameter~$L_0$.

We wish to consider in Eq.~\eqref{eq:conf-chi-F} the long interaction duration limit, $\Delta\to\infty$. While $\eta_{\text{mid}}$ has dropped out of the last expression in Eq.~\eqref{eq:conf-chi-F}, we recall that $\eta<0$ by construction, so the limit must be understood so that $\eta_{\text{mid}}\to-\infty$ as $\Delta \to\infty$ in a way that keeps $\eta_{\text{mid}}+\Delta$ negative. We also note that because the single-mode contribution to ${(\eta \eta')}^{-1} W(\eta,\eta')$ depends on $\eta$ and $\eta'$ only through the difference $\eta-\eta'$, the coupled system is effectively stationary, and 
$\mathcal{F}^{\chi}_{k_0}(E)$ grows at large $\Delta$ proportionally to the total interaction time~$\Delta$. We therefore consider the transition probability per unit time, or  transition rate, given by 
\begin{align}
\dot{\mathcal{F}}^{\chi}_{k_0, \Delta}(E) 
& = 
\frac{H^2}{2 k_0 L_0} 
\frac{1}{2\Delta} \Bigg|  \int_{-\Delta}^{\Delta} d\sigma \, \cos^4\!\left( \frac{\pi \sigma}{2\Delta}\right)\, e^{-i\left( E + k_0 \right) \sigma} \Bigg|^2
\notag\\
& = \frac{9 \pi^8 H^2 \Delta}{4 k_0 L_0} 
R\bigl(\Delta(E+k_0)\bigr) , 
\label{eq:conf-chi-Fdot}
\end{align}
where 
\begin{align}
R(x) = \frac{\sin^2 \! x }{x^2 {(x^2 - \pi^2)}^2 {(x^2 - 4 \pi^2)}^2} , 
\label{eq:Rfunc-def}
\end{align}
and 
Eq.~\eqref{eq:Rfunc-def} is understood in the limiting sense at the zeroes of the denominator. Note that $R$ is smooth and it has the falloff $R(x) = O(x^{-10})$ as $x\to\pm\infty$. 
For any smooth and rapidly decreasing test function~$\Upsilon$, we therefore have 
\begin{align}
\Delta \int_{-\infty}^{\infty} dy \, R(\Delta y) \Upsilon(y) 
= 
\int_{-\infty}^{\infty} dx \, R(x) \Upsilon(x/\Delta) 
\xrightarrow[\Delta \to \infty]{}
\Upsilon(0)
\int_{-\infty}^{\infty} dx \, R(x) 
= 
\frac{35}{288 \pi^7} \Upsilon(0) , 
\end{align}
using dominated convergence to take the limit under the integral. 
In the limit of long interaction duration, $\Delta \to\infty$, 
Eq.~\eqref{eq:conf-chi-Fdot} hence gives 
\begin{align}
\dot{\mathcal{F}}^{\chi}_{k_0, \infty}(E) 
= 
\lim_{\Delta \to \infty}
\dot{\mathcal{F}}^{\chi}_{k_0, \Delta}(E) 
& = 
\frac{35 \pi H^2}{128 k_0 L_0} 
\delta(E + k_0) . 
\label{eq:conf-chi-Fdot-longtime}
\end{align}
The transition rate is a Dirac delta peak for a de-excitation at the frequency of the mode to which the detector couples, $E = - k_0$. 

We conclude that the conformal time microphysics single-mode detector, with the novel coupling to the field, reacts very similarly to a single-mode detector in Minkowski spacetime where the spatial sections have been compactified to a torus. In the latter case, the response function of a static detector reads 
\begin{equation}
    \mathcal{F}^{\phi}_{\mathscr{M}}(E) = \frac{1}{2 L_0^3} \sum_{\vec{k}} \Bigg| \int_{t_{\text{mid}}-\Delta}^{t_{\text{mid}}+ \Delta}  \frac{dt }{\sqrt[4]{k^2+m^2} } \cos^4 \! \left( \frac{\pi ( t - t_{\text{mid}} )}{2 \Delta} \right) e^{-iEt} e^{- i t \sqrt{k^2+m^2}} \Bigg|^2,
\end{equation} 
where 
$\vec{k} = 2\pi \vec{n}/L_0$, $\vec{n} = (n,l,r) \in \mathbb{Z}^3$, $k = \sqrt{\vec{k}\cdot \vec{k}}$, and $m>0$ is the mass of the scalar field. For a detector coupled to only one mode, with $k = k_0>0$, proceeding as in Eqs.~\eqref{eq:conf-chi-F}--\eqref{eq:conf-chi-Fdot-longtime} gives the long interaction time transition rate 
\begin{align}
\dot{\mathcal{F}}^{\phi}_{\mathscr{M},\infty}(E)
= 
\frac{35 \pi}{128 \sqrt{k_0^2 + m^2} L_0^3} 
{\textstyle{\delta \! \left( E + \sqrt{k_0^2 + m^2} \right)}} , 
\end{align}
with a Dirac delta de-excitation peak at the energy of the mode to which the detector couples, $E = -\sqrt{k_0^2 + m^2}$. 

Physically, this similarity indicates that the novel detector is closely adapted to the expanding background\footnote{Taking the definition of unitary implementable dynamics presented in Ref. \cite{Triada} one could ask to what extent this preservation of the form of the response function, and thus of the detector's information, is related to a preservation of coherence.}.

\subsection{Standard coupling}

For the standard coupling, starting from Eq.~\eqref{eq:conftime-F-phi} and proceeding as above gives the response function
\begin{align}
\mathcal{F}^{\phi}_{k_0}(E) 
&= 
\frac{H^2}{2 k_0 L_0^3} \Bigg| \int_{-\Delta}^{\Delta} d\sigma \, \cos^4 \! \left( \frac{\pi \sigma}{2\Delta}\right) (\sigma+\eta_{\text{mid}}) \, e^{-i\left( E+  k_0\right) \sigma} \Bigg|^2 , 
\label{eq:conf-phi-F}
\end{align}
where we recall that $\eta_{\text{mid}} + \Delta <0$. Since $\eta_{\text{mid}}$ appears in Eq.~\eqref{eq:conf-phi-F}, the response is not stationary. The notion of a long interaction duration is hence subtle, especially as this limit must have $\eta_{\text{mid}}\to-\infty$ when $\Delta \to\infty$ so that  $\eta_{\text{mid}}+\Delta$ stays negative. 

As an example, we consider the long time limit in which $\eta_{\text{mid}} = -\Delta$ and $\Delta\to\infty$. This means that the detector is turned off in the asymptotic future, $\eta\to0_-$, quartically in~$\eta$. From Eq.~\eqref{eq:conf-phi-F} we obtain 
\begin{align}
\frac{1}{\Delta^3}\mathcal{F}^{\phi}_{k_0}(E) 
&= 
\frac{H^2 \Delta}{k_0 L_0^3} S\bigl(\Delta(E+k_0)\bigr) , 
\label{eq:conf-phi-F-explicit}
\end{align}
where $S$ is an elementary function, expressible in terms of rational and trigonometric functions. From the explicit formula for $S$, which we do not reproduce here, it can be verified that $S$ is smooth, it has the falloff $S(x) = O(x^{-10})$ as $x\to\pm\infty$, and 
$\int_{-\infty}^{\infty} dx \, S(x) = 
35(96 \pi^2 - 205)/(4608 \pi)$. 
It follows, as in Eq.~\eqref{eq:conf-chi-Fdot-longtime}, that 
\begin{align}
\frac{1}{\Delta^3}\mathcal{F}^{\phi}_{k_0}(E) 
\xrightarrow[\Delta \rightarrow \infty]{}
\frac{35(96 \pi^2 - 205)}{4608 \pi}
\frac{H^2 }{k_0 L_0^3} \delta (E+k_0) . 
\label{eq:conf-phi-F-longtime}
\end{align}

The factor $\delta (E+k_0)$ on the right-hand side of Eq.~\eqref{eq:conf-phi-F-longtime} is, as expected, showing a Dirac delta peak for a de-excitation at the frequency of the mode to which the detector couples. The factor $\Delta^{-3}$ on the left-hand side of Eq.~\eqref{eq:conf-phi-F-longtime} shows, however, that the detector's transition probability diverges proportionally to 
$\Delta^3$ as $\Delta\to\infty$, 
and the transition rate diverges proportionally to~$\Delta^2$. 

We conclude that the conformal time microphysics single-mode detector, with the standard coupling to the field, 
is more divergent in the long-time limit than the detector with the novel coupling to the field, giving both a diverging de-excitation probability and a diverging de-excitation rate. 

In Sec.\ \ref{sec:discrete}
we shall introduce an idealised switching function that will shed light on the differences between the standard coupling and the novel coupling for the conformal time microphysics single-mode detector. \\

\section{$\mathbb{R}^3$ spatial topology}
\label{sec:noncompact}

In the previous sections, we have addressed a scalar field in locally de Sitter spacetime with compactified spatial sections. Several of the characteristics observed in the detector's response are related to the finite spatial volume. In this section, we consider $\mathbb{R}^3$ spatial sections, for both the standard coupling \eqref{Conventional} and the novel coupling~\eqref{Interact}.

\subsection{Wightman function}

With $\mathbb{R}^3$ spatial topology, our metric \eqref{eq:conf-desittemetric} covers half of de Sitter spacetime, as discussed in Sec.~\ref{subsec:compactification}. 
The Bunch-Davies state that results from the adiabatic criterion \eqref{eq:adiabatic-form} is defined on all of de Sitter spacetime: it is alternatively known as the Euclidean vacuum or the Chernikov-Tagirov vacuum, and it is invariant under all continuous isometries of de Sitter spacetime~\cite{ChernikovTagirov,BAllen,Mukhanov,Bunch:1978yq}. As our choice of the field mass, $m=H\sqrt{2}$, makes the field equivalent to a massless conformal field, the Wightman function in the Bunch-Davies state is given by \cite{Bunch:1978yq,BD}
\begin{equation} \label{TheStandard}
    W^{\phi}_{\text{dS}}(\eta,\vec{x}; \eta',\vec{x}') = - \frac{H^2}{4\pi^2} \frac{\eta \eta'}{(\eta - \eta' -i \varepsilon)^2 - |\vec{x}-\vec{x}'|^2}, 
\end{equation}
where the distributional limit $\epsilon\to0_+$ is understood. 

As a consistency check, we note that $W^{\phi}_{\text{dS}}$ 
in Eq.~\eqref{TheStandard} can be obtained as the $L_0\to\infty$ limit of $W^{\phi}_{\modes}(\eta,\vec{x}; \eta',\vec{x}')$, given in Eq. ~\eqref{ConvWeight}. 
Interpreting Eq.~\eqref{ConvWeight} as a Riemann sum, with the cell volume ${(2\pi/L_0)}^3$, we have 
\begin{align}
W^{\phi}_{\modes}(\eta,\vec{x}; \eta',\vec{x}') 
&=
\frac{H^2 \ \eta \eta'}{16 \pi^3} 
\left(\frac{2\pi}{L_0}\right)^{\! 3}
\sideset{}{'}\sum_{\vec{k}} 
\frac{1}{k} e^{i k \left(|\eta|- |\eta'|\right)}
    \, e^{ i \vec{k} \cdot (\vec{x}-\vec{x}')} 
\notag\\
&\xrightarrow[L_0\to\infty]{}
\frac{H^2 \ \eta \eta'}{16 \pi^3} 
\int \frac {d^3 \vec{k}}{k}
e^{i k \left(|\eta|- |\eta'|\right)}
    \, e^{ i \vec{k} \cdot (\vec{x}-\vec{x}')}
\notag\\
& = W^{\phi}_{\text{dS}}(\eta,\vec{x}; \eta',\vec{x}') , 
\label{eq:Wphiosc-to-Wphi-ds}
\end{align}
where the last step can be verified as in Minkowski spacetime, by inserting the convergence factor $e^{-\epsilon k}$ and performing the elementary integrals, and finally recalling that $\eta$ and $\eta'$ are negative. Note that the limit in Eq.~\eqref{eq:Wphiosc-to-Wphi-ds} comes solely from the finite-$L_0$ oscillator modes, with no contribution from the corresponding zero mode. 

We emphasise that $W^{\phi}_{\text{dS}}$ is invariant under all continuous isometries of de Sitter spacetime, despite this not being transparent from expression \eqref{TheStandard} in the spatially flat foliation. In particular, given a timelike geodesic, $W^{\phi}_{\text{dS}}$ is invariant under time translations in the static coordinate patch centred on this geodesic, and hence invariant under time translations along the geodesic. We shall use this property in our analysis.

\subsection{Standard coupling}

Consider a detector with the standard coupling, given by the $L_0\to\infty$ limit of Eq.~\eqref{Conventional}. 

The pull-back of $W^{\phi}_{\text{dS}}$ in Eq.~\eqref{TheStandard} to the inertial trajectory $(t,x,y,z) = (t, 0, 0, 0)$ reads, using Eq.~\eqref{eq:eta-vs-t-and-a}, 
\begin{align}
W^{\phi}_{\text{dS}}(t,t') = 
- \frac{H^2}{16\pi^2} \frac{1}{\sinh^2 \! \bigl(\frac12 H (t-t' - i \epsilon) \bigr)} , 
\label{eq:ds-wight-pullback}
\end{align}
in which the stationarity is apparent because the dependence on $t$ and $t'$ is only through the combination $t-t'$. The Fourier transform of Eq.~\eqref{eq:ds-wight-pullback} in $t-t'$ is 
\begin{align}
\widehat W^{\phi}_{\text{dS}}(E) &= 
\int_{-\infty}^{\infty} dt \, e^{-iEt} \, 
W^{\phi}_{\text{dS}}(t, 0)
\notag \\
&= \frac{1}{2\pi} \frac{E}{e^{2\pi E/H}-1} . 
\label{ds-phi-longtime-response}
\end{align}
The response of an inertial detector in the long-time limit therefore follows the Planckian spectrum in the de Sitter temperature 
$T_{\text{dS}} = H/(2\pi)$ ~\cite{Bunch:1978yq,Gibbons:1977mu,BD}.

For the finite time detector with the switching profile $\Theta$~\eqref{switching}, we now have 
\begin{align} 
\mathcal{F}_{\text{dS}}^\phi(E) 
&= 
\int dt \, dt' \, 
\Theta(t) \Theta(t') 
\, W^{\phi}_{\text{dS}}(t,t') . 
\label{eq:F-dS}
\end{align}
Using the convolution theorem, this can be written as \cite{UnruhWaiting}
\begin{align} 
\mathcal{F}_{\text{dS}}^\phi(E) 
& = \frac{1}{2\pi} \int_{-\infty}^{\infty} d \omega\, {\bigl|\widehat\Theta(\omega)\bigr|}^2 \, \widehat W^{\phi}_{\text{dS}} (\omega + E), 
\end{align}
where \cite{LoukoMann}
\begin{equation} 
\widehat\Theta(\omega) = 
\frac{3 \sin(\Delta\omega)}{\omega \left( 1-\left(\frac{\Delta\omega}{\pi}\right)^2\right)
\left( 4-\left(\frac{\Delta\omega}{\pi} \right)^2\right)} , 
\label{eq:Thetahat-formula}
\end{equation}
and we have set $t_{\text{mid}}=0$ without loss of generality, by the time translation invariance. 
From 
Eqs.~\eqref{ds-phi-longtime-response}--\eqref{eq:Thetahat-formula}, we obtain
\begin{equation}\label{Rphi}
\mathcal{F}^{\phi}_{\text{dS}} (E)= \frac{9}{4\pi^2} \int_{-\infty}^{\infty} dz \, 
\frac{z + 2\pi E / H}{e^{z + 2\pi E / H} -1}
\; \frac{\sin^2 \! \left( \frac{\Delta H}{2\pi} \,z \right)}{z^2 \left( 1-\left(\frac{\Delta H}{2\pi^2} \,z \right)^2\right)^2  \left( 4-\left(\frac{\Delta H}{2\pi^2} \,z\right)^2\right)^2} , 
\end{equation}
changing variables by $\omega = Hz/(2\pi)$. Note that $z$ is dimensionless.\\

\subsection{Novel coupling} 

Consider then a detector with the novel coupling. For finite~$L_0$, this coupling is given by Eq.~\eqref{Interact}. In the $L_0\to\infty$ limit, we define the novel coupling by including the time-dependent factor $H a(t) = e^{Ht}$ that appears in Eq.~\eqref{Interact}. From Eq.~\eqref{eq:F-dS}, we then have the novel response function 
\begin{align}
    \mathcal{F}^{\chi}_{\text{dS}} (E)
    &= \int dt \, dt' \, e^{Ht}\Theta(t)\, e^{Ht'}\Theta(t') \, e^{-iE(t -t')} \,W^{\phi}_{\text{dS}}(t, t') . 
\end{align}
The effect of the novel coupling is hence to replace the switching function $\Theta$
\eqref{switching} by the new switching function $\underline{\Theta}(\tau) = \Theta(\tau) e^{H\tau}$. Proceeding as in Eqs.~\eqref{eq:F-dS}--\eqref{Rphi}, we find 
\begin{align} 
\mathcal{F}_{\text{dS}}^\chi(E) 
& = \frac{1}{2\pi} \int_{-\infty}^{\infty} d \omega\, {\bigl|\widehat{\underline\Theta}(\omega)\bigr|}^2 \, \widehat W^{\phi}_{\text{dS}} (\omega + E), 
\label{eq:ds-F-chi-gen}
\end{align}
where 
\begin{equation}
\widehat{\underline{\Theta}}(\omega) =   \frac{3 \sinh \bigl( \Delta (H-i\omega)\bigr) }{(H-i\omega) \left(1+ \frac{\Delta^2}{\pi^2}{(H-i\omega)}^2\right)\left(4+ \frac{\Delta^2}{\pi^2}{(H-i\omega)}^2\right)}. 
\label{eq:undertheta-fourier}
\end{equation}
We have again set $t_{\text{mid}}=0$ without loss of generality: the value of $t_{\text{mid}}$ affects only the overall constant in the response, and as the spatial volume is now infinite, this overall constant has no geometric significance, and it can be absorbed into the coupling constant in the definition of the novel coupling. 

From 
Eqs.~\eqref{eq:ds-F-chi-gen}
and~\eqref{eq:undertheta-fourier}, we now have 
\begin{align}
\mathcal{F}^{\chi}_{\text{dS}} (E)
& = 
\frac{9}{4\pi^2} \int_{-\infty}^{\infty} dz\, 
\frac{z + 2\pi E/H}{e^{z + 2\pi E/H} - 1} \;
\frac{\sinh^2 ( \Delta H) + \sin^2 \! \left( \frac{\Delta H}{2 \pi} \, z \right)}{4\pi^2 +z^2}
\notag\\[2ex]
& \hspace{3ex}
\times
\frac{1}{\displaystyle\left(
\left(\frac{\Delta^2H^2}{\pi^2}
\left(\frac{z^2}{4\pi^2} + 1\right) -1 \right)^2 + \frac{4\Delta^2H^2}{\pi^2}
\right)
\left(
\left(\frac{\Delta^2H^2}{\pi^2}
\left(\frac{z^2}{4\pi^2} + 1\right) -4 \right)^2 + \frac{16\Delta^2H^2}{\pi^2}
\right)} , 
\label{Rchi} 
\end{align}
changing variables by $\omega = Hz/(2\pi)$. 
Note that $z$ is again dimensionless.\\

\subsection{Numerical results}

Figures \ref{fig:NCA}--\ref{fig:UCBB} in Appendix \ref{app:numerical} show numerical plots 
for $\mathcal{F}^{\phi}_{\text{dS}}$ \eqref{Rphi} and $\mathcal{F}^{\chi}_{\text{dS}}$~\eqref{Rchi}, as a function of the dimensionless gap $E/H$ and the dimensionless interaction duration~$\Delta H$. The plots suggest that both $\mathcal{F}^{\phi}_{\text{dS}}$ and $\mathcal{F}^{\chi}_{\text{dS}}$ increase linearly at large negative~$E/H$: we verify this increase analytically in Appendix~\eqref{app:dS-as}, and we show that the increase is faster for~$\mathcal{F}^{\chi}_{\text{dS}}$. There is no evidence of de-excitation resonances similar to those that arose for finite $L_0$ in Secs.~\ref{sec:UDWdetector} and~\eqref{sec:singlemodedetector}.

\section{Discrete switches and the intuitive notion of a particle detector}
\label{sec:discrete}

As noticed in the previous sections, some of the effects that are being observed in the comparison between the two detector models are related to the intrinsic differences between their respective Wightman functions. The objective of the present section is to explore in further detail the effects of the different couplings on their respective response functions. We start by revisiting the intuitive notion underlying the operation of a particle detector. The two-level system, when its trajectory is fixed and while it is \textit{switched on}, is expected to \textit{beep\/} when the system is (de-)excited. Ideally, we want to avoid introducing any external input to the system when turning on and off the interactions \cite{FamousSatz, Satz, Switchings}. With this aim, when permitted, e.g. by the spacetime dimensions, instantaneous switching functions are commonly employed \cite{TovLou, Adam}. In a similar vein, we are motivated to consider instantaneous switching functions, placed equidistantly but nevertheless arbitrarily apart, while restricting our analysis to a single-mode coupling, thereby preventing divergences in the response functions. Hence, we obtain as straight as possible a relation of both Wightman functions to the respective (de-)excitation probabilities of the detector. It is important to mention, however, that the resulting response functions, instead of being interpreted by themselves, are intended to exploit their differences and thus to be contrasted with one another. 
A complementary approach, with switching functions that consist of Dirac delta peaks, has been considered in Ref.~\cite{Polo-Gomez:2023gaz}.

\subsection{Discrete unit-switching functions}

\begin{figure}[h!] \label{SWI}
    \centering
    \includegraphics[width=0.5\textwidth]{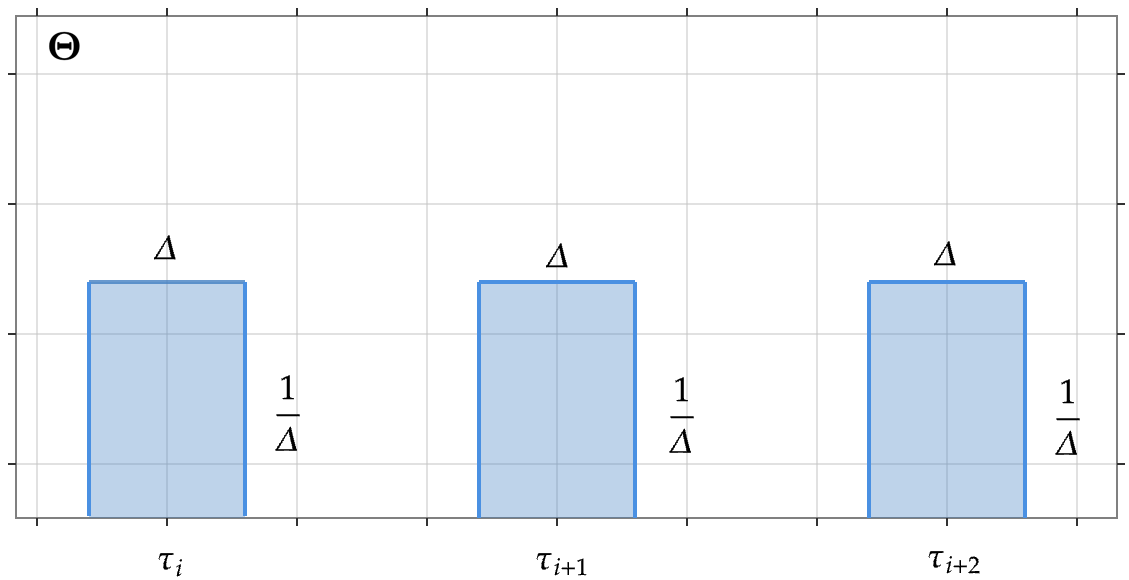}
    \caption{Unit-switching functions, defined as rectangular functions of unit area centred at times $\tau_j$.}
    \label{fig:USF}
\end{figure}

Let us consider the following instantaneous switching functions with a unit area. We introduce what we term the \textit{unit-switching functions\/}, denoted by $\mathbf{\Theta}_{\tau_j}$, to refer to rectangular (or \textit{Heaviside Pi\/}, or \textit{gate\/}) functions with a height equal to~$1/\Delta$, a width equal to $\Delta$ and with midpoint~$\tau_j$, as shown in Fig.~\ref{fig:USF}. We shall study detector switching functions that are given by the $N$-sum of such unit-switching functions, i.e.\ $\Theta(\tau) = \sum_{j=0}^{N} \mathbf{\Theta}_{\tau_j}(\tau)$. 

It is clear from Fig.\ \ref{fig:USF} that if we let $\Delta$ run sufficiently small, the switching function $\Theta(\tau)$ becomes an $N$-sum over~$\tau_j$ and each summand plays the role of a Dirac delta peaked at the corresponding value of~$\tau_j$. 

For the novel coupling, the fact that we are only considering a finite sum lets us then write the following response functions, for a single-mode characterised by $k_0$:
\begin{equation}
    \mathcal{F}^{\chi}_{\modes} (E) = \frac{1}{2 k_0 L_0}  \Bigg| \sum_{j=0}^{N}  e^{-i\frac{E}{H}\tau_j} e^{i k_0 \exp \left( -\tau_j \right) }\Bigg|^2.
\label{eq:unit-Fchiosc1}
\end{equation}
Let us consider the unit-switching functions to be equidistant. Specifically, $\tau_j = -j \ln M - \ln (2\pi/k_0)$, where $M \in \mathbb{Z}^{+}\setminus \{ 1\}$ is arbitrary. Then, it is clear that the second exponential in Eq.~\eqref{eq:unit-Fchiosc1} is equal to one. Hence, we obtain
\begin{equation}
    \mathcal{F}^{\chi}_{\modes}(E) = \frac{1}{2 k_0 L_0}  \Bigg| \left(\frac{2\pi}{k_0} \right)^{i\frac{E}{H}}\sum_{j=0}^{N}  \left(M^{j} \right)^{i\frac{E}{H}}\Bigg|^2= {\frac{1}{2 k_0 L_0}}  \Bigg| \sum_{j=0}^{N}  \left( M^{j} \right)^{i\frac{E}{H}}\Bigg|^2, \quad \text{for} \,\, M=2,3,4,\hdots
\label{eq:unit-Fchiosc2}
    \end{equation}
We can compute the sum in Eq.~\eqref{eq:unit-Fchiosc2} as that of a geometric series, getting $(1-M^{i(N+1)E/H})/(1-M^{iE/H})$. Taking its square modulus, writing the resulting expression as exponentials and using the identity $1-\cos x= 2\sin^2 (x/2)$, we obtain
\begin{equation}
    \mathcal{F}^{\chi}_{\modes} (E) = \frac{1}{2 k_0 L_0} \Bigg| \frac{\sin \! \left(\,(N+1) \frac{E}{2H} \,\ln{ M }\right)}{\sin \! \left( \frac{E}{2H} \,\ln{M}\right)}\Bigg|^2 = \frac{1}{2 k_0 L_0} \ U_N^2 \! \left[\cos \! \left( \frac{E}{2H}\, \ln{ M} \right) \right] ,
\end{equation}
where $U_N\left[\cos  \theta  \right]= \sin ( (N+1) \theta)/\sin \theta$ is the $N$-th Chebyshev polynomial of the second kind. From the first equality, we can see that, by considering values of energy gaps sufficiently small, such that overall $(N+1) E \ln M /(2H)$ remains small, the response function is approximately $N+1$, the number of on/off-switches, as expected for a detector probing particles. Furthermore, in the second equality, we note that the Chebyshev polynomial has a (cyclic) extremum in the interval $[-1,1]$ and, at it, $U_N[1]=N+1$. More generally, the obtained response function contains the interference of a sum of phases inside of a square modulus. We can select an energy gap $E=4\pi Hp/\ln M$, for any $p \in \mathbb{Z}$, for which the unit-switching functions have no interference with one another. For these particular cases, the response function does not depend on the choice of the position of the unit-switches, but only on the number of on/off-switches, in accord with the underlying intuitive concept of a detector. Similar considerations to those above apply to the other extremum of $U_N$ at~$-1$.   

Repeating the above analysis for the standard coupling, we find
\begin{equation}
    \mathcal{F}^{\phi}_{\modes} = \frac{2\pi^2}{ k_0^3 L_0^3} \Bigg| \sum_{j=0}^{N}  \left(M^{j}  \right)^{1+i\frac{E}{H}}\Bigg|^2, \quad \text{for} \,\, M=2,3,4,\hdots 
    \end{equation}
Then, we conclude that
\begin{equation}
    \mathcal{F}^{\phi}_{\modes} = \frac{2\pi^2 }{ k_0^3 L_0^3} \; \frac{\left(1-M^{N+1}\right)^2 + 4 M^{N+1} \sin^2 \! \left( \frac{E}{2H} \ln{M} \right) U_N^2 \! \left[\cos \left( \frac{E}{2H}\, \ln{ M} \right) \right]}{\left( 1-M\right)^2 + 4 M \sin^2 \! \left( \frac{E}{2H}\ln{M} \right) }.
\end{equation}
Note that the response function for the standard coupling has a different dependence on the mode $k_0$ and is independent of the compactification parameter $L_0$, given that the product $k_0 L_0$ is fixed by the labels $n_0$, $l_0$ and $r_0$. Moreover and more importantly, considering again that there is no interference between the unit-switching functions, i.e. choosing again an energy gap $E=4\pi Hp/\ln M$, we obtain a response function proportional to $(1-M^{N+1})^2/(1-M)^2$. Thus, the response function depends explicitly on the choice of the constant~$M$, namely the instant from which we arrange the unit-switches, even when no interference between the unit-switching functions is contemplated. 

These types of detectors help us observe significant differences between the conventional and novel couplings. 
Furthermore, we can actually adopt a concept of unitary evolution as the one introduced in Ref.~\cite{Triada}. Then, a possible interpretation of these observed differences could be attributed to the unitarity, or the lack thereof, of the specific dynamics under investigation. In this sense, it becomes apparent that, while the standard coupling is adapted to a dynamical evolution influenced by the nonstationary background, the novel coupling is adapted to a field dynamics in which the notion of a particle seems to persist during the detection interval. Moreover, we would like to emphasise that the time dependence of the novel coupling comes entirely from its background dependence. Thus, a priori, no knowledge of the evolution is needed other than the information that can be measured directly from the background.

By an analogous analysis, we obtain similar results when considering a detector governed by an evolution in conformal time. 
For the novel coupling, with unit-switching functions centred equidistantly at $\eta_j = j 2\pi/k_0$, we find 
\begin{equation}
 \mathcal{F}^{\chi}_{\modes}(E) = \frac{H^2}{2 k_0 L_0}  \Bigg| \sum_{j=0}^{N} \exp \left(  -i  \frac{2\pi E}{k_0} j \right)  \Bigg|^2.
\label{eq:Fchiosc-cilinder1}
\end{equation}
Evaluating the sum in Eq.~\eqref{eq:Fchiosc-cilinder1} as a geometric series gives 
\begin{equation}
    \mathcal{F}^{\chi}_{\modes}(E) = \frac{H^2}{2 k_0 L_0} \ U_N^2 \! \left[ \cos \! \left( \frac{\pi E}{k_0}\right) \right].
\end{equation}
Just as in the case of a particle detector coupled in cosmological time, we can choose an energy gap $E= 2 pk_0$, for $p \in \mathbb{Z}$, such that there is no interference between the unit-switching functions. The response function is thus proportional to $N+1$, the number of on/off-switches. 

For the standard coupling, we find 
\begin{equation}
    \mathcal{F}^{\phi}_{\modes}(E) = \frac{2\pi^2 H^2}{ k_0^3 L_0^3} \Bigg| \sum_{j=0}^{N}  j \exp \left(  -i  \frac{2\pi E}{k_0} j \right) \Bigg|^2.
\label{eq:Fphiosc-cilinder1}
\end{equation}
Let us denote the sum in Eq.~\eqref{eq:Fphiosc-cilinder1} by~$S_{N+1}$, and write $b=2\pi E/k_0$. Note that 
\begin{equation}
S_{N+1}-e^{-ib}S_{N+1} = -1 - N e^{-ib(N+1)}+\sum^{N}_{j=0} e^{-ibj} .
\end{equation} 
Therefore, we obtain
\begin{equation}
    \mathcal{F}^{\phi}_{\modes}(E) = \frac{2\pi^2 H^2}{ k_0^3 L_0^3} \Bigg| \frac{\left(1-e^{-ib(N+1)}\right) +\left(N e^{-ib(N+1)}+1\right)\left(e^{-ib}-1\right)}{\left( e^{-ib}-1\right)^2}\Bigg|^2.
\end{equation}
Computing the square modulus and simplifying, we get
\begin{equation}
\mathcal{F}^{\phi}_{\modes}(E) = \frac{\pi^2 H^2}{k_0^3 L_0^3} \,  \frac{1 + N + N^2 - N (1 + N) \cos b- (1 + N) \cos(b N) + N \cos\left(b (1 + N)\right)}{4 \sin^4 \! \left(\frac{b}{2}\right) }.
\label{eq:Fphiosc-cilinder2}
\end{equation}

Interestingly, the limit of Eq.~\eqref{eq:Fphiosc-cilinder2} for energy gaps that represent no interference between the unit-switching functions ($E= 2 pk_0$, for $p \in \mathbb{Z}$) is equal to $ N^2 ( N + 1 )^2 /4$ (times the proportionality constant of the front) 
for the case $p=0$ (implying E=0), and is indeterminate for any other integer value of~$p$. Hence, we see that none of these cases possesses the relation with the number of on/off-switches expected for a sequence of accumulated detections of particles.

\section{Conclusions} \label{sec:conclusions}

In the analysis of particle detectors within the framework of first-order perturbation theory, it is interesting to explore the existing freedom to use various switching functions, different numbers of energy gaps, alternative parametrisations of the detector time evolution, vary the physical size of the detector or consider distinct trajectories. Among all the possibilities allowed by this freedom, an intriguing one is to allow a background-dependent coupling of the detector with the field in a nonstationary spacetime. This enables us to investigate, as in the present context of expanding cosmologies, different field dynamics. Indeed, through a background-dependent (and thus time-dependent) scaling in the coupling, a detector is able to disregard dynamical excitations that are only due to the background variation and then present a response function specially adapted to scalar field dynamics disentangled from the background.

The particle detectors considered in this paper (referred to as standard and novel) display inherited differences in their outcomes from one another. Numerical results suggest that the novel coupling presents a more concentrated response function around approximately the field mode momentum in contrast with the standard one, which is observed to be more dispersed. Furthermore, the analysis of both detectors (coupled to one mode only), evolving in cosmological and conformal time for long intervals of interaction, exhibits for the standard case a complete spreading of the signal response function (or a divergent transition rate for the case considered in Sec. \ref{SecVB}), growing unboundedly for large negative energy gaps, while the novel coupling is capable of preserving the peak form in its response function and transition amplitude. It is worth mentioning that the latter is a phenomenon that is not only present in the studied scenario, but also occurs when considering the transition rate of a detector in (compact) Minkowski spacetime. Surprisingly, we have determined that the existence of a resonance is a characteristic feature of the de Sitter background with compact spatial sections, compared to the noncompact de Sitter case where no resonance is found, and a linear growth for large negative energy gaps is proven by an asymptotic analysis. Lastly, by introducing a sequence of gate switching functions with an area equal to the unity and equidistant from each other, we have demonstrated that, with a single operative mode, there exists a direct relation between the response functions and their respective Wightman functions. Interestingly, by choosing these switching functions to be essentially instantaneous and with no mutual interference, we obtain that the response function with the novel coupling is proportional to the number of on/off-switches, as one would intuitively expect for a cumulative series of detections. On the contrary, in the standard case, even with the same choice of switching functions, no similar relation is obtained regardless of the specific time associated with the energy levels of the detector (i.e. cosmological or conformal time). Indeed, the response function, in this case, depends on the choice of the separation of the basic switching functions of unit area, a fact that lets us think that the background spacetime, or strictly speaking its evolution, is interfering with the readings on the detector, even for static trajectories.

The analysis and results presented in this paper open a window to discuss to what extent decoherence plays a role in the standard detector signal. In this sense, the novel detector is well-adapted to probe the part of the field dynamics that is unitarily implementable, according to the notion of unitarity studied in Ref.~\cite{Triada}. Moreover, this work paves the way to explore detectors with similar background-dependent couplings in more general cosmological scenarios, such as the Bianchi I and the Kantowski-Sachs spacetime, which are homogeneous but anisotropic. Certainly, both spacetimes have significant physical relevance and can ultimately provide substantial intuition in quantum theories of gravity, such as the case of Loop Quantum Cosmology. Indeed, Bianchi I cosmologies are related to singularity resolution and Big Bounce-related issues (see e.g.\ Ref. \cite{AshtekarWilsonEwing, MMP1, MMP2}). On the other hand, 
the interior geometry of a nonrotating black hole has a description as a Kantowski-Sachs spacetime \cite{Morales, AshtekarBojowald, CorMorRu, CamGamPu}.
Demanding the invariance of the vacuum under the spatial symmetries and the adoption of dynamics that are unitarily implementable in the aforementioned sense\footnote{These conditions were put forward in Ref. \cite{Homogen} for isotropic cosmologies, and in Refs. \cite{JGJV, JGJV2} for more general ultrastatic spacetimes.}, it has been proven that one can in fact select a unique family of unitary equivalent Fock representations, both in Bianchi I  \cite{DreamTeam} and Kantowski-Sachs \cite{GunsNRoses, Beatles}. One can use the same strategy in these cases and reassign part of the field dynamics to the background. 
In this regard, an appealing possibility would be to extend our analysis of background-dependent couplings to these scenarios and, in particular, analyse couplings adapted to the specific notion of particle associated with a recent proposal of non-oscillating vacuum with asymptotically diagonal Hamiltonian dynamics, comparing the results with others corresponding to better studied black hole vacua~\cite{Wald}.

\section*{Acknowledgments}
A.T.-C. is grateful to Antonio Vicente-Becerril and Bruno Falcone for constructive discussions on numerics and to the Gravity Laboratory Group of
the University of Nottingham for the interesting discussions on the subject. 
We thank an anonymous referee for helpful presentational suggestions. 
G.A.M.M.\ acknowledges
financial support from Project No.\ PID2020-118159GB-C41 and Project No.\ PID2023-149018NB-C41, both funded by MCIN/AEI/10.13039/501100011033/, and Project No.\ 2024AEP005 from the Spanish National Research Council (CSIC). The work of J.L.\ was supported by the United Kingdom Research and Innovation Science and Technology Facilities Council [grant numbers ST/S002227/1, ST/T006900/1 and ST/Y004523/1].
For the purpose of open access, the authors have applied a CC BY public copyright licence to any Author Accepted Manuscript version arising. 

\appendix

\section{Numerical results}
\label{app:numerical}

This appendix collects numerical plots of the response functions studied in the main text. To avoid cluttering, the graphs indicate all response functions by the symbol~$F$, while the captions specify which response function is plotted.\\

\begin{figure}[h!]
    \centering
    \begin{minipage}{0.405\textwidth} 
        \centering
        \includegraphics[width=\linewidth]{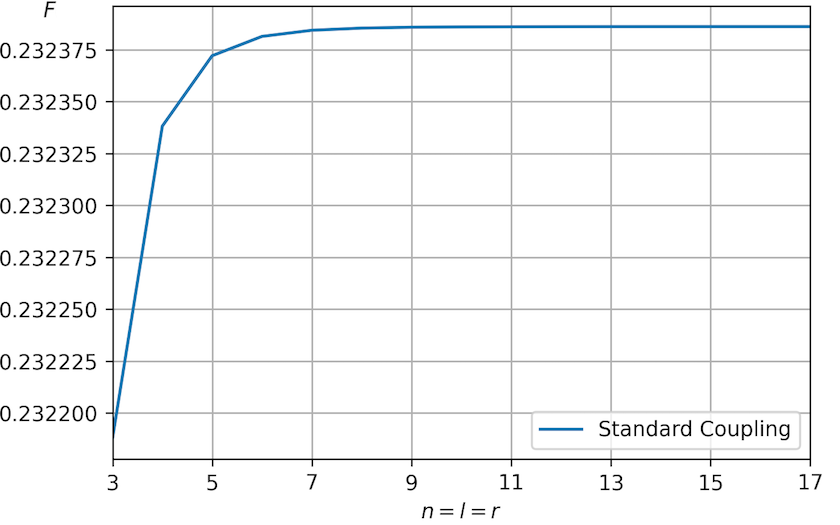}
        \caption{Standard coupling response function $\mathcal{F}^{\phi}_{\modes}(E)$ \eqref{resp} for $E/H=-3$, $L_0=0.5$ and $H\Delta = 3.5$, with the sum over $\vec{k}$ truncated as explained in the main text. The quantity on the horizontal axis is the maximum value of $|n|$, $|l|$ and $|r|$ included in the sum.}
        \label{fig:ConvA}
    \end{minipage}\hfill
    \begin{minipage}{0.405\textwidth} 
        \centering
        \includegraphics[width=\linewidth]{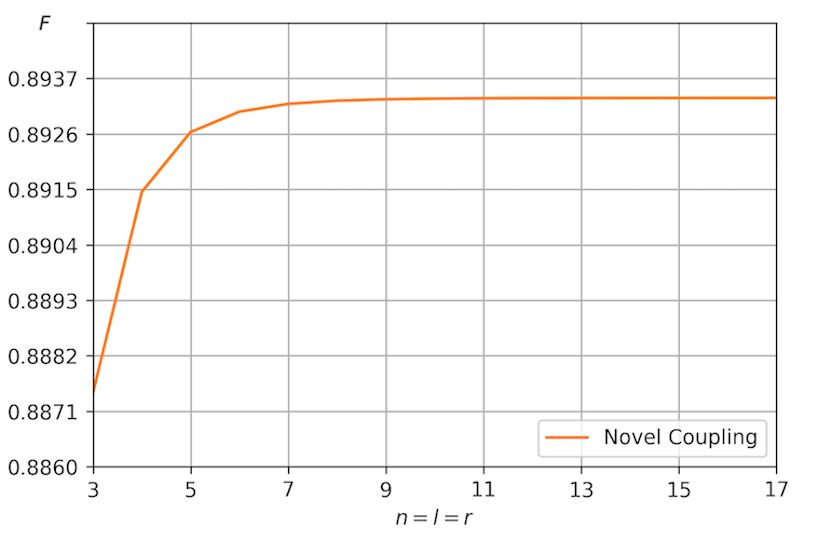}
        \caption{As in Fig. \ref{fig:ConvA} but for the novel coupling response function $\mathcal{F}^{\chi}_{\modes}(E)$~\eqref{chiresponse}.}
        \label{fig:ConvB}
    \end{minipage}\hfill
    \centering
    \begin{minipage}{0.5\textwidth} 
        \centering
        \includegraphics[width=\linewidth]{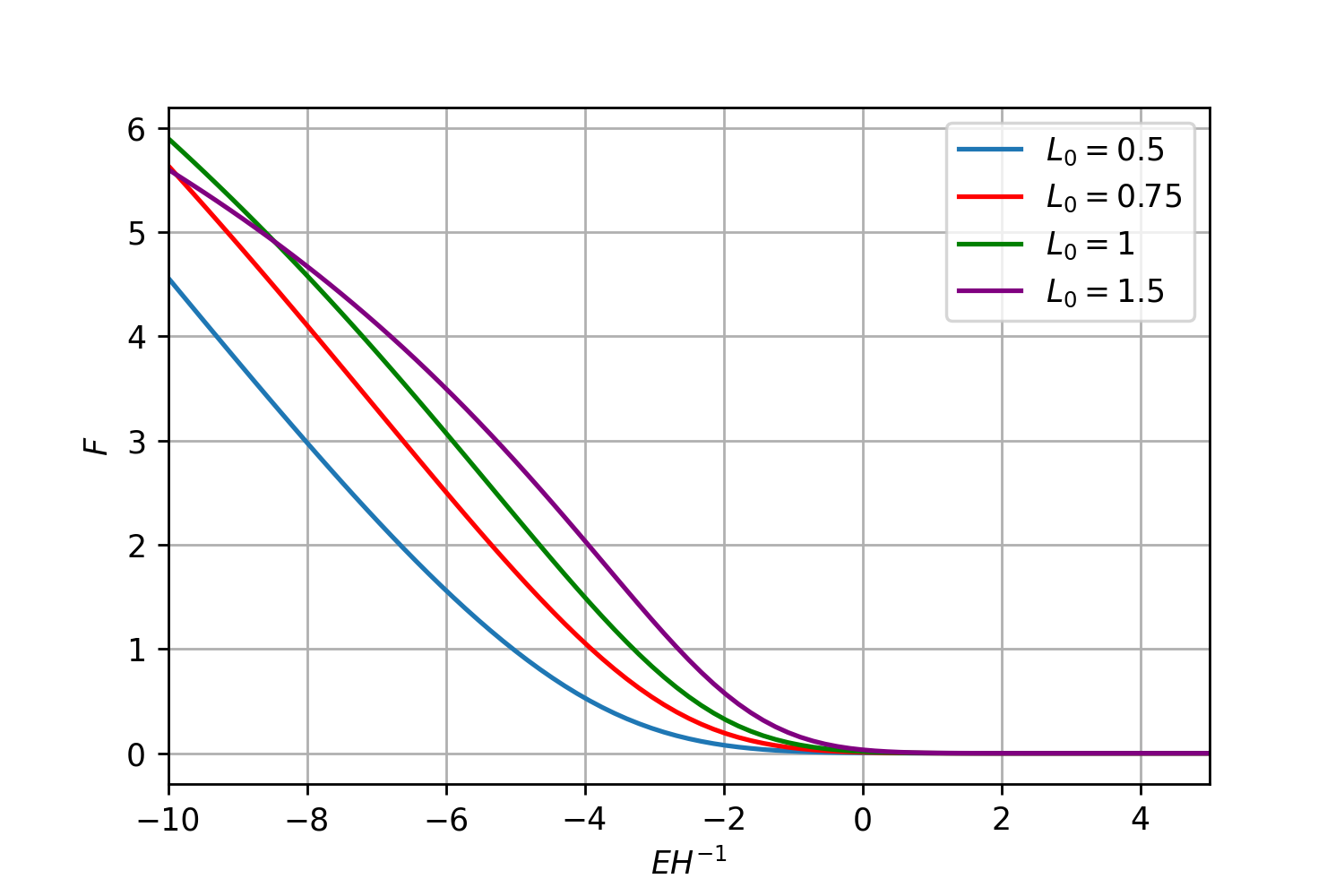}
        \caption{Standard coupling response function $\mathcal{F}^{\phi}_{\modes}(E)$ \eqref{resp} as a function of the dimensionless gap~$E/H$, for $H\Delta = 3.5$ and selected values of~$L_0$. The numerical evaluation truncates $|n|$, $|l|$ and $|r|$ at~11.}
        \label{fig:EvolA}
    \end{minipage}\hfill 
    \begin{minipage}{0.5\textwidth} 
        \centering
        \includegraphics[width=\linewidth]{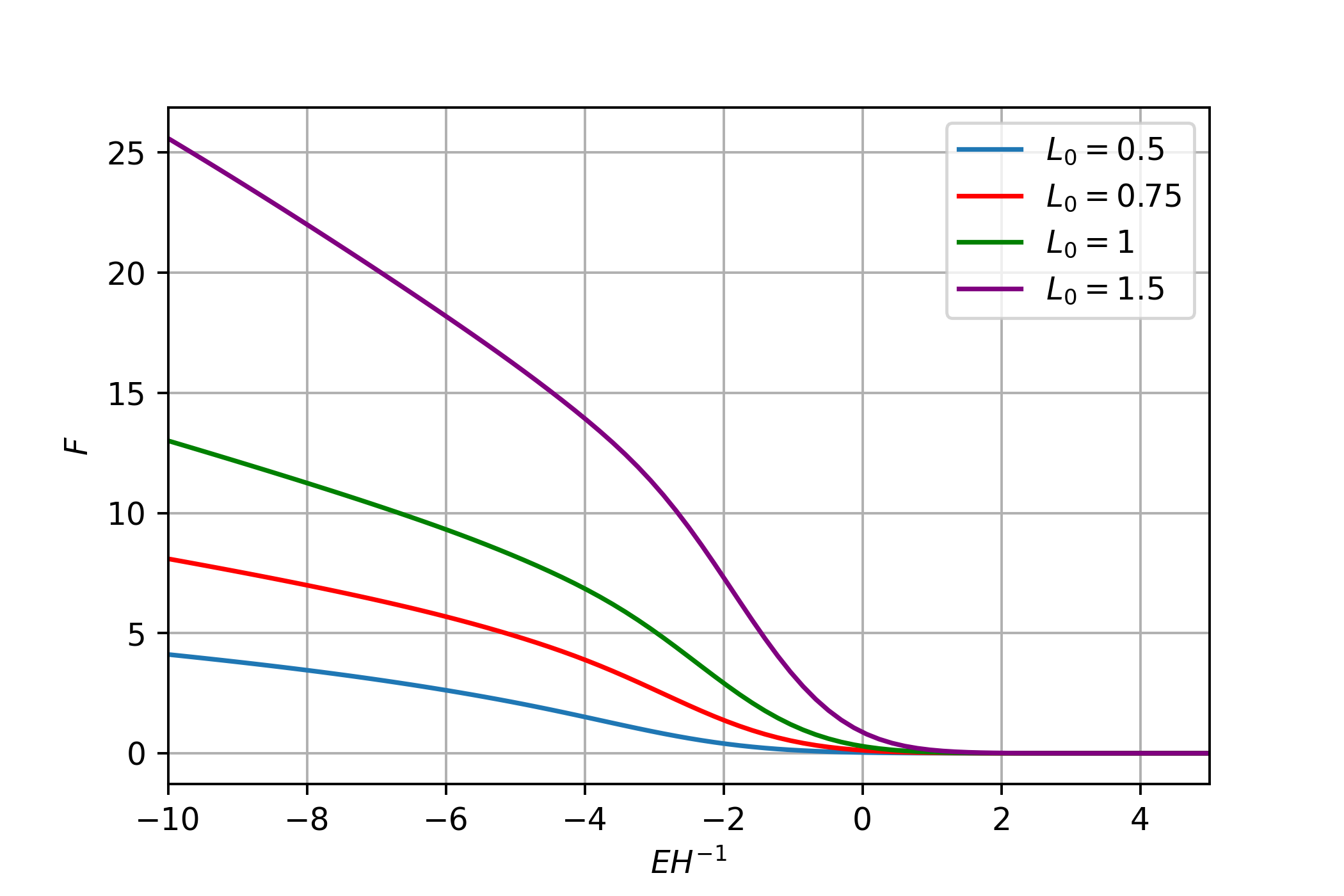}
        \caption{As in Fig. \ref{fig:EvolA} but for the novel coupling response function $\mathcal{F}^{\chi}_{\modes}(E)$~\eqref{chiresponse}. 
        The numerical evaluation truncates $|n|$, $|l|$ and $|r|$ at~20.}
        \label{fig:EvolB}
    \end{minipage}\hfill 
\end{figure}

\begin{figure}[h!]
    \centering
    \begin{minipage}{0.46\textwidth} 
        \centering
        \includegraphics[width=\linewidth]{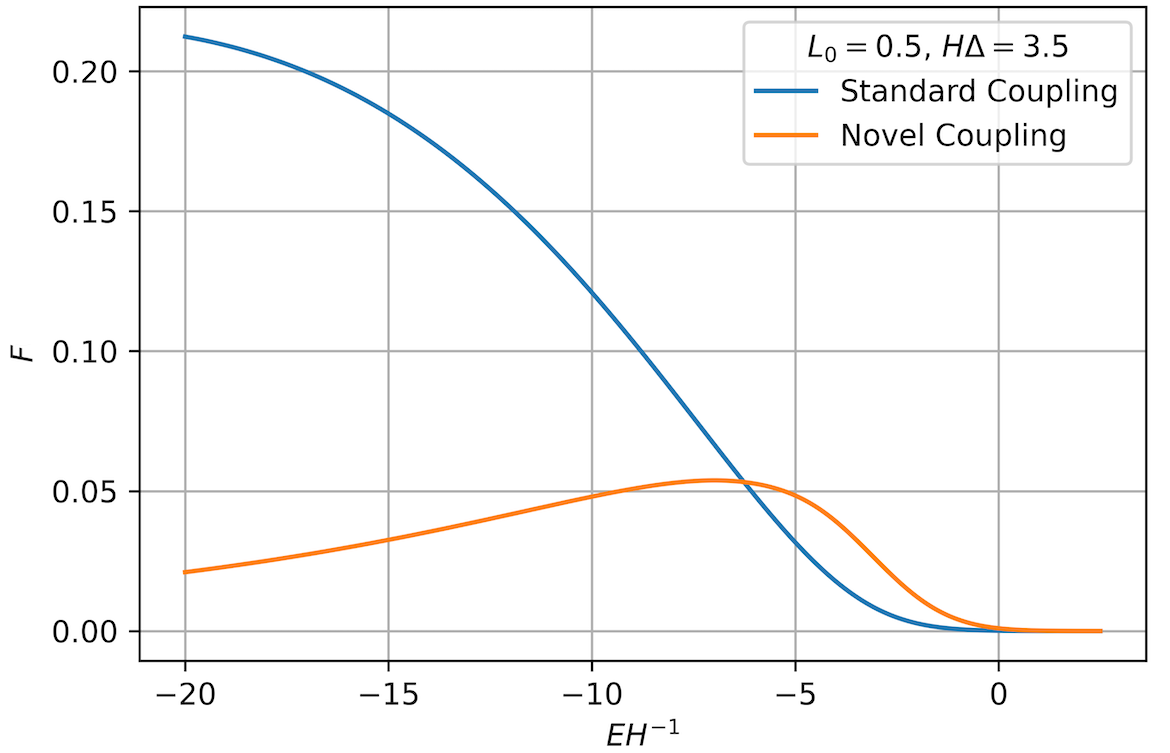}
        \caption{Single-mode standard coupling response function $\mathcal{F}^{\phi}_{k_0}(E)$ \eqref{resp-singlemode} and single-mode novel coupling response function $\mathcal{F}^{\chi}_{k_0}(E)$ \eqref{chiresponse-singlemode} as a function of the dimensionless gap~$E/H$, for $H\Delta=3.5$, $k_0 = 2\pi/L_0$ and $L_0=0.5$.}
        \label{fig:OneModeA}
    \end{minipage}\hfill
    \begin{minipage}{0.46\textwidth} 
        \centering
        \includegraphics[width=\linewidth]{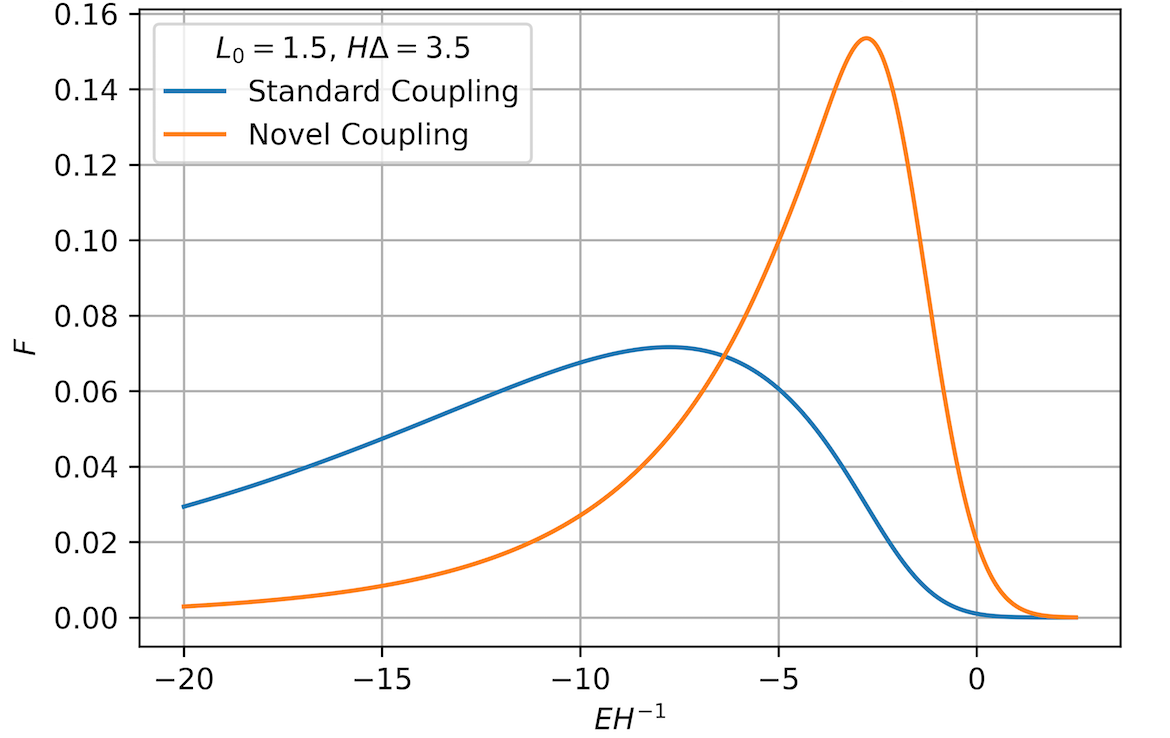}
        \caption{As in Fig. \ref{fig:OneModeA} but with $L_0=1.5$.}
        \label{fig:OneModeB}
    \end{minipage} \hfill \\ \vspace{.3cm} 
     \centering

    \begin{minipage}{0.47\textwidth} 
        \centering
        \includegraphics[width=\linewidth]{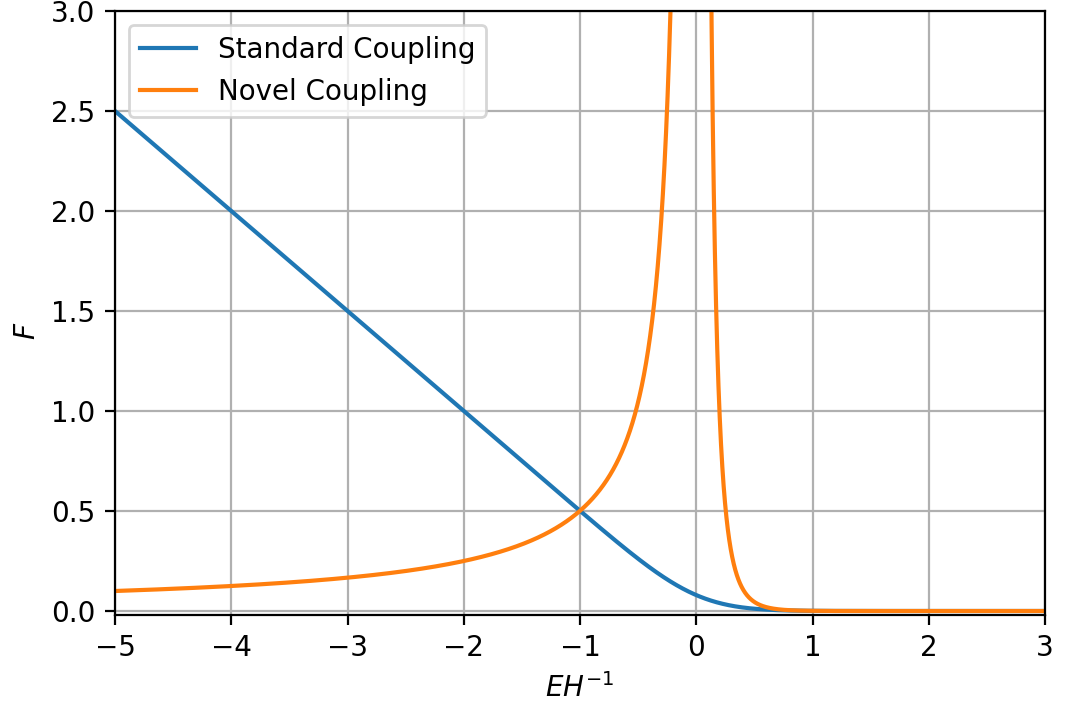}
        \caption{Long-time limits \eqref{eq:novelcoupling-longtime} and \eqref{eq:standardcoupling-longtime} of the single-mode novel coupling and standard coupling response functions
        as a function of the dimensionless gap~$E/H$, 
        for $k_0 = 2\pi/L_0$ and $L_0=1$.}
        \label{fig:Delta}
    \end{minipage}\hfill
    \\
    \end{figure}

    \begin{figure}[h!]
    \begin{minipage}{0.48\textwidth} 
        \centering
        \includegraphics[width=\linewidth]{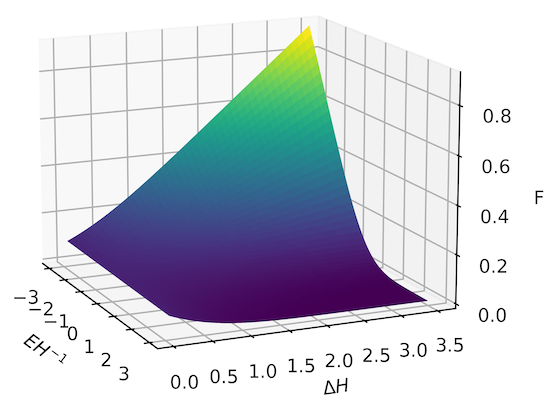}
        \caption{Standard coupling response function $\mathcal{F}^{\phi}_{\text{dS}}$ \eqref{Rphi} in full de Sitter spacetime, as a function of the dimensionless gap $E/H$ and the dimensionless interaction duration $\Delta H$.}
        \label{fig:NCA}
    \end{minipage}\hfill
    \begin{minipage}{0.48\textwidth} 
        \centering
        \includegraphics[width=\linewidth]{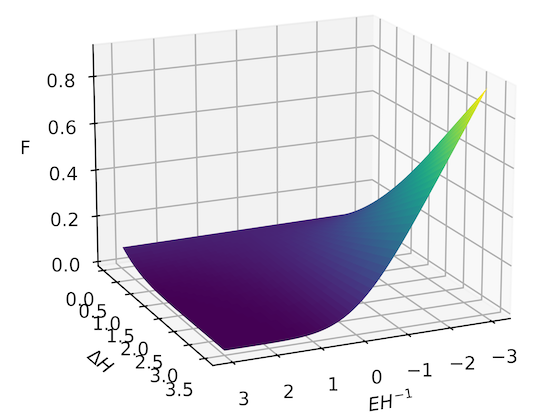}
        \caption{As in Fig. \ref{fig:NCA} but from a different perspective.}
        \label{fig:STBB}
    \end{minipage}
    \centering
    \begin{minipage}{0.48\textwidth} 
        \centering
        \includegraphics[width=\linewidth]{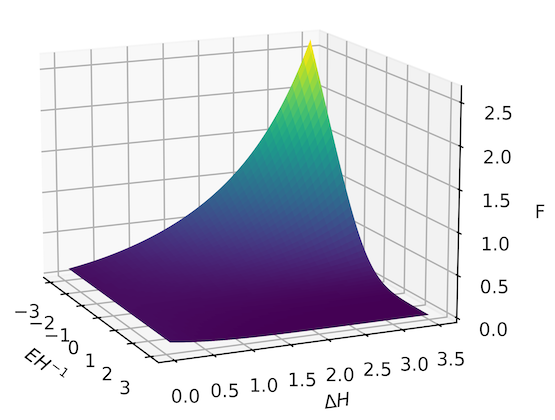}
        \caption{As in Fig. \ref{fig:NCA} but for the novel coupling response function $\mathcal{F}^{\chi}_{\text{dS}}$ \eqref{Rchi}.}
        \label{fig:NCB}
    \end{minipage}\hfill
    \begin{minipage}{0.48\textwidth} 
        \centering
        \includegraphics[width=\linewidth]{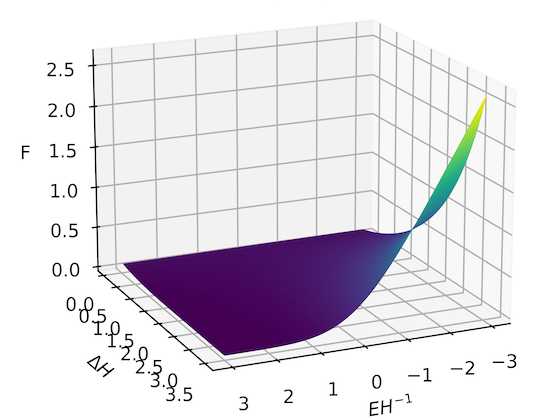}
        \caption{As in Fig. \ref{fig:NCB} but from a different perspective.}
        \label{fig:UCBB}
    \end{minipage}
\end{figure}

\clearpage

\section{$\mathcal{F}^{\phi}_{\text{dS}}$ and $\mathcal{F}^{\chi}_{\text{dS}}$ at large negative $E/H$}
\label{app:dS-as}

In this appendix we verify that $\mathcal{F}^{\phi}_{\text{dS}}$ \eqref{Rphi} and $\mathcal{F}^{\chi}_{\text{dS}}$ \eqref{Rchi} have linear growth at large negative~$E/H$. 

We write $\mathcal{F}^{\phi}_{\text{dS}}$ and $\mathcal{F}^{\chi}_{\text{dS}}$  
as 
\begin{subequations}
\begin{align}
\mathcal{F}^{i}_{\text{dS}}(E) &= \frac{9}{4\pi^2} G^i(2\pi E/H) , 
\label{eq:LFDs-formula}
\\
G^i(\alpha) &= 
\int_{-\infty}^{\infty} dz \, g(z + \alpha) \, h_i(z) , 
\label{eq:LFDs-integral}
\\
g(u) &= \frac{u}{e^u -1}, 
\end{align}
\end{subequations}
where $i \in \{\phi,\chi\}$, and the expressions for $h_i(z)$ may be read off from 
Eqs.~\eqref{Rphi} and~\eqref{Rchi}. $g$~and $h_i$ are smooth, $g$ and $h_\chi$ are positive, $h_\phi$ is non-negative, and $h_i$ are even with $h_i(z) = O\bigl(z^{-10}\bigr)$ as $z\to\pm\infty$. 
The large negative $(E/H)$-behaviour of $\mathcal{F}^{i}_{\text{dS}}(E)$ is hence determined by the large negative $\alpha$-behaviour of~$G^i(\alpha)$. 

Suppressing for the moment the index~$i$, we break the integration domain in Eq.~\eqref{eq:LFDs-integral} into $(-\infty, -\alpha]$ and $[-\alpha,\infty)$ and decompose 
$G$ as 
\begin{subequations}
\begin{align}
G(\alpha) &= G_1(\alpha) + G_2(\alpha) + G_3(\alpha) , 
\\
G_1(\alpha) &= 
- \int_{-\infty}^{-\alpha} dz \, (z+\alpha) h(z) , 
\\
G_2(\alpha) &= 
\int_{-\infty}^{-\alpha} dz \, 
g(-z-\alpha) \, h(z) , 
\\
G_3(\alpha) &= 
\int_{-\alpha}^{\infty} dz \, g(z+\alpha) \, h(z) , 
\end{align}
\end{subequations}
where $G_1$ and $G_2$ arise from $(-\infty, -\alpha]$ using the identity $g(u) = - u + g(-u)$, and $G_3$ corresponds to $[-\alpha,\infty)$. 

For~$G_1$, we have 
$G_1(\alpha) = 
- \int_{-\infty}^{\infty} dz \, z\, h(z) - \alpha \int_{-\infty}^{\infty} dz \, h(z) + \int_{-\alpha}^{\infty} dz \, z\, h(z) + \alpha \int_{-\alpha}^{\infty} dz \, h(z)= - \alpha \int_{-\infty}^{\infty} dz \, h(z) + O \bigl( \alpha^{-8}\bigr)$ as $\alpha \to -\infty$, 
where the first integral vanishes since $z \,h(z)$ is odd, and we have used the $O\bigl(z^{-10}\bigr)$ falloff of $h(z)$ to estimate the last two integrals. 

To estimate~$G_2$, we note that by the smoothness and falloff of $h$ there exist positive constants $M_1$ and $M_2$ such that $0\le h(z) \le M_1$ for all $z$ and $0\le h(z) \le M_2/z^{10}$ for $z\ge1$. We also note the inequality $0 < g(u) \le  (1+u) e^{-u}$ for $u\ge0$. When $\alpha < -2$, we hence have 
\begin{align}
0 < G_2(\alpha) & \le  
\int_{-\infty}^{-\alpha} dz \,  (1 - z - \alpha) e^{z+\alpha} \, h(z) 
\notag \\
& = \int_{-\infty}^{-\alpha/2} dz \,  (1 - z - \alpha) e^{z+\alpha} \, h(z) 
+ \int_{-\alpha/2}^{-\alpha} dz \,  (1 - z - \alpha) e^{z+\alpha} \, h(z) 
\notag\\
& \le M_1 \int_{-\infty}^{-\alpha/2} dz \,  (1 - z - \alpha) e^{z+\alpha}
+ M_2 \int_{-\alpha/2}^{-\alpha} dz \,  \frac{1 - z - \alpha}{z^{10}} \,  e^{z+\alpha} . 
\end{align}
The integral multiplying $M_1$ evaluates to $\bigl( 2 - \tfrac12 \alpha \bigr) e^{\alpha/2}$, which is $O \bigl(\alpha e^{\alpha/2}\bigr)$ as $\alpha \to -\infty$. For the integral multiplying~$M_2$, elementary estimates give the bound $\bigl( 1 - \tfrac12 \alpha \bigr) {(2/\alpha)}^{10}$, which is $O\bigl(\alpha^{-9}\bigr)$ as $\alpha \to -\infty$. Hence $G_2(\alpha) = O\bigl(\alpha^{-9}\bigr)$ as $\alpha \to -\infty$. 

For~$G_3$, since $0 < g(u) \le1$ for $u\ge0$, we have 
$
0 < G_3(\alpha) \leq \int_{-\alpha}^\infty dz \, h(z) = O \bigl( \alpha^{-9}\bigr)$ 
as $\alpha \to -\infty$, where the last equality comes from the $O\bigl(z^{-10}\bigr)$ falloff of~$h(z)$. 

Collecting everything, we have $G(\alpha) = 
- \alpha \int_{-\infty}^{\infty} dz \, h(z) + O\bigl(\alpha^{-8}\bigr)$ as $\alpha \to -\infty$. 

From Eq.~\eqref{eq:LFDs-formula} we then have, restoring the index~$i$, 
\begin{align}
\mathcal{F}^{i}_{\text{dS}}(E) = 
- \frac{9E}{2\pi H}
\int_{-\infty}^{\infty} dz \, h_i(z) + O\bigl(E^{-8}\bigr) , 
\label{eq:Fi-appendix-finalas}
\end{align}
as $E/H \to -\infty$. We do not include the elementary expressions for $\int_{-\infty}^{\infty} dz \, h_i(z)$, but it may be proved that $\int_{-\infty}^{\infty} dz \, h_\chi(z) > \int_{-\infty}^{\infty} dz \, h_\phi(z)$. 
This shows that the novel coupling de-excitation response grows at large negative $E/H$ faster than the standard coupling de-excitation response, for all~$\Delta H$.

\clearpage 
\end{document}